\documentclass[reprint,
superscriptaddress,
 amsmath,amssymb, 
 aps, prb, longbibliography]{revtex4-2}
\usepackage{todonotes}
\usepackage{siunitx}
\usepackage{multirow}

\usepackage{listings}
\usepackage{xcolor}

\definecolor{blue}{HTML}{1f77b4}
\definecolor{orange}{HTML}{ff7f0e}
\definecolor{green}{HTML}{2ca02c}
\definecolor{red}{HTML}{d62728}
\definecolor{purple}{HTML}{9467bd}
\definecolor{brown}{HTML}{8c564b}
\definecolor{pink}{HTML}{e377c2}
\definecolor{gray}{HTML}{7f7f7f}
\definecolor{olive}{HTML}{bcbd22}
\definecolor{cyan}{HTML}{17becf}

\usepackage{graphicx}
\usepackage[utf8]{inputenc} 
\usepackage[T1]{fontenc}
\usepackage{amsmath}
\usepackage{color}
\usepackage[pdfusetitle]{hyperref}
\usepackage{mathtools}

\usepackage{etoolbox}
\apptocmd{\sloppy}{\hbadness 4000\relax}{}{} 
\hypersetup{
  colorlinks,
  citecolor=blue,
  linkcolor=blue,
  urlcolor=black}

\newcommand{\kkk}{\mathbf{k}}

\newcommand{\bbb}{\mathbf{b}}

\newcommand{\Ge}{Mn$_3$Ge}
\newcommand{\Sn}{Mn$_3$Sn}

\begin{document}

\title{Ground state magnetic structure of \texorpdfstring{\Sn{}}{Mn3Sn}}
\author{Jeppe Jon Cederholm} \thanks{These authors contributed equally to this work.}
\affiliation{Institut Laue-Langevin, 71 avenue des Martyrs, CS 20156, 38042 Grenoble cedex 9, France} 
\affiliation{Laboratory for Quantum Magnetism, Institute of Physics, \'{E}cole Polytechnique F\'{e}d\'{e}rale de Lausanne (EPFL), CH-1015 Lausanne, Switzerland}
\affiliation{Nanoscience Center, Niels Bohr Institute, University of Copenhagen, 2100 Copenhagen, Denmark}

\author{Zhian Xu} \thanks{These authors contributed equally to this work.}
\affiliation{State Key Laboratory of Quantum Functional Materials, School of Physical Science and Technology, ShanghaiTech University, Shanghai 201210, China}

\author{Yanfeng Guo} \affiliation{State Key Laboratory of Quantum Functional Materials, School of Physical Science and Technology, ShanghaiTech University, Shanghai 201210, China}

\author{Martin Ovesen} \affiliation{Computational Atomic-scale Materials Design, Department of Physics, Technical University of Denmark, DK-2800 Kongens Lyngby, Denmark}

\author{Thomas Olsen} \affiliation{Computational Atomic-scale Materials Design, Department of Physics, Technical University of Denmark, DK-2800 Kongens Lyngby, Denmark}

\author{Kristine M. L. Krighaar} \affiliation{Nanoscience Center, Niels Bohr Institute, University of Copenhagen, 2100 Copenhagen, Denmark}

\author{Chrystalla Knekna} \affiliation{Zernike Institute for Advanced Materials, University of Groningen, 9747 AG Groningen, Netherlands}
\affiliation{Nanoscience Center, Niels Bohr Institute, University of Copenhagen, 2100 Copenhagen, Denmark}

\author{Jian Rui Soh} \affiliation{Quantum Innovation Centre (Q.InC), Agency for Science, Technology and Research (A*STAR), 2 Fusionopolis Way, Innovis \#08-03, Singapore 138634, Singapore}
\affiliation{Centre for Quantum Technologies, National University of Singapore, 3 Science Drive 2, Singapore 117543, Singapore}

\author{Youngro Lee} \affiliation{Laboratory for Quantum Magnetism, Institute of Physics, \'{E}cole Polytechnique F\'{e}d\'{e}rale de Lausanne (EPFL), CH-1015 Lausanne, Switzerland}

\author{Navid Qureshi} \affiliation{Institut Laue-Langevin, 71 avenue des Martyrs, CS 20156, 38042 Grenoble cedex 9, France} 

\author{Jose Alberto Rodriguez Velamazan} \affiliation{Institut Laue-Langevin, 71 avenue des Martyrs, CS 20156, 38042 Grenoble cedex 9, France} 

\author{Eric Ressouche} \affiliation{Univ. Grenoble Alpes, CEA, IRIG, MEM, MDN,  38000 Grenoble, France}

\author{Andrew T. Boothroyd} \affiliation{Department of Physics, University of Oxford, Clarendon Laboratory, Parks Road, Oxford OX1 3PU, United Kingdom}

\author{Henrik Jacobsen} \email{henrik.jacobsen.fys@gmail.com}
\affiliation{Data Management and Software Centre, Asmussens Allé 305, 2800 Kongens Lyngby, Denmark}
\affiliation{Nanoscience Center, Niels Bohr Institute, University of Copenhagen, 2100 Copenhagen, Denmark}

\begin{abstract}
We use spherical neutron polarimetry to determine the ground state magnetic structure of \Sn{}.  We find that \Sn{}  adopts an inverse triangular structure with spins parallel to $\langle100 \rangle$ (Type III)  rather than spins parallel to $\langle 110 \rangle$ (Type IV). Density functional theory calculations reveal no energy difference between these two structures, suggesting that the selection is caused by subtle effects such as sixth-order anisotropy.
Partial control of the magnetic domain population through a moderate magnetic field is key to distinguish between the two models. We find that three of the six domains are approximately equally populated, while the others have negligible population. 
Upon entering the low temperature incommensurate phase, the domain structure is lost. The domains in this phase are decoupled from the magnetic field, and can therefore not be controlled by any known method.  
\end{abstract}

\maketitle

\section{Introduction}
The non-collinear antiferromagnets Mn$_3$\textit{X} (\textit{X} = Sn, Ge, Ga) display a very large anomalous Hall effect (AHE) at room temperature \cite{nakatsuji2015large,kiyohara2016giant,nayak2016large,Sung2018,Raju2024,Chen2021a}. The sign of the AHE can be switched by the application of a small magnetic field, by strain \cite{Dasgupta2022,Ikhlas2022}, and even with electrical fields, which makes these materials interesting candidates for antiferromagnetic spintronic applications \cite{Baltz2018,DalDin2024,Zheng2024}. 
The AHE in Mn$_3$\textit{X} arises from a large Berry curvature, which originates from Weyl points near the Fermi level \cite{chen2014anomalous,nakatsuji2015large,Li2023}. For such Weyl points to exist, either time reversal or inversion symmetry must be broken \cite{Boothroyd2022}. In  Mn$_3$\textit{X}, time reversal symmetry is broken by the non-collinear magnetic structure.

\begin{figure}
    \centering
    \includegraphics[width=1\linewidth]{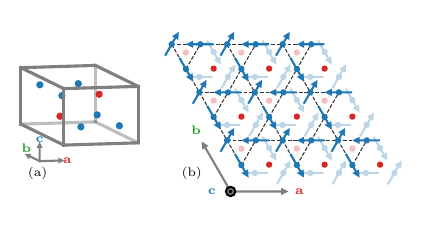}
    \caption{The crystal structure of \Sn{}, with Mn atoms marked in blue and Sn atoms marked in red. Faded atoms are at a lower layer than non-faded atoms. (a)  The unit cell of \Sn{}. (b) Several unit cells viewed along the $c$ axis, showing how each layer of Mn atoms forms a kagome lattice. }
    \label{fig:Fig1}
\end{figure}

The Mn atoms form a kagome lattice, as illustrated in Fig.~\ref{fig:Fig1}. Non-collinear antiferromagnetic order with propagation vector $\kkk=0$ sets in below the Néel temperature, which for \Sn{} is 420 K. Here,  neighboring Mn spins within the kagome layers align at $120^\circ$ to each other accompanied by a tiny ferromagnetic moment. 
Depending on the exact composition \cite{Ikhlas2020,Park2025}, \Sn{} undergoes a transition to  an incommensurate (IC) phase upon cooling below $T_S\approx 290$ K, with two propagation vectors $\kkk_L\sim (0,0,0.08)$ and $\kkk_T\sim (0,0,0.11)$   \cite{Kouvel1964, Zimmer1972, Cable1993,Song2020,Chen2024,wang_flat_2023}. 
The IC phase restores time reversal symmetry \cite{wang_flat_2023}, and the AHE goes to zero. 

\Sn{} crystallizes in space group $P6_3/mmc$, and several magnetic structures  with moments at 120$^{\circ}$ and a small ferromagnetic component are compatible with this space group, as discussed in Refs.~\cite{pj1990determination,Zimmer1972,Tomiyoshi1986,soh2020ground}. Some of the structures are illustrated in Fig.~\ref{fig:Fig2}~\footnote{Note that models (III) and (IV) were mislabelled as Pc'mm' and Pcm'm', respectively, in the original article by P.J Brown et al. \cite{pj1990determination}}.
Several neutron diffraction experiments have been performed to determine which of these structures is correct in the high temperature phase \cite{Tomiyoshi1986,Nagamiya1982,pj1990determination,Cable1993,wang_flat_2023}. Despite this effort, the structure has not yet been uniquely determined; it has been narrowed down to two similar structures \cite{pj1990determination}. Following Ref.~\cite{soh2020ground}, we label these structures type III or IV, as shown in Fig.~\ref{fig:Fig2}. A recent study in magnetic fields found type III to be the most likely ground state \cite{Chen2024}. However, no study has determined the ground state in zero field. 

Part of the difficulty comes from the fact that both type III and IV allow for six magnetic domains, illustrated in Fig.~\ref{Fig:D3DataAndFit}, which complicates the analysis of neutron diffraction data.   The domain population can strongly influence the behavior of the material. For example, the switching of the sign of the AHE by a magnetic field is because the magnetic domains (partially) align with the field.
Understanding and controlling the domains is therefore vital to determine the magnetic structure of \Sn{}, which in turn is 
an important step towards applications in spintronics \cite{DalDin2024}.

Spherical neutron polarimetry (SNP) on \Ge{} has shown that its magnetic structure is type IV \cite{soh2020ground}. Given the similar structures and transport behavior, it has therefore been assumed that \Sn{} and \Ge{} have the same magnetic structure, e.g., Refs. \cite{Song2020, Chen2021a, Higo2022}.

Here, we use SNP to show that the magnetic structure of \Sn{} is not the same as that of \Ge{}, but actually type III as also found in the in-field study \cite{Chen2024}. Interestingly, our density functional theory (DFT) calculations show no difference between the band structure of type III and IV, indicating that the energy difference between the two types of order must be  small and most likely governed by sixth order anisotropy.  The population of the six domains in the high temperature phase can be controlled to some extent via an external magnetic field. We furthermore show that the domain distribution in the low-temperature IC phase is decoupled from the magnetic field. Finally, we use unpolarized neutron diffraction to determine the temperature dependence of the incommensurate structure below $T_S$.

\section{Experimental details}
High-quality \Sn{} single crystals were synthesized via a Bi-flux method. Stoichiometric mixtures of Mn powder (99.95\% purity), Sn granules (99.999\% purity), and Bi granules (99.999\% purity) were combined in a molar ratio of 3:1:3. The mixture was sealed in an alumina crucible under a partial argon atmosphere within a quartz ampoule and heated to 1150 $^\circ$C over 15 hours. After holding at this temperature for 10 hours to ensure homogeneity, the assembly was slowly cooled to 700 $^\circ$C at a controlled rate of 50 $^\circ$C/$h$ and annealed for an additional 10 hours. To optimize crystal growth, the temperature was further reduced to 300 $^\circ$C at a slower rate of 2 $^\circ$C/$h$ and maintained for 50 hours. Finally, the ampoule was removed from the furnace, and the \Sn{} crystals were isolated from the residual Bi flux via centrifugation. The sample used here was a $3\times0.8\times0.8$ mm$^3$ single crystal. 

Magnetic susceptibility and standard transport measurements confirm the large AHE at 300 K, and show a magnetic phase transition at 292~K. Details of these measurements can be found in the appendix. 

Spherical neutron polarimetry (SNP) measurements were performed on D3 at the Institut Laue--Langevin (ILL) in Grenoble, France \cite{Jacobsen2023}. The SNP method involves measuring the diffraction intensity of spin-up and spin-down polarized neutrons for all 9 combinations of initial and final neutron polarization along three orthogonal directions, $x$, $y$ and $z$. On D3, the Blume-Maleev coordinate system is used \cite{Andrews_book}, in which  $x$ is along the scattering vector $\textbf{Q}$, $z$ is perpendicular to the scattering plane, and $y$ completes a right-handed coordinate system. The final polarization is measured along these three directions,  resulting in a $3 \times 3$ polarization matrix. The matrix elements are defined as $P_{fi}$, where $i$ and $f$ refer to the initial and final polarization directions ($x,y,z$). The experiment was performed in zero field using Cryopad ~\cite{Tasset1999,Lelievre2005}. The incident wavelength was $\lambda=0.83$~\AA{}.  
The incident beam was polarized by Bragg reflection from a monochromator of ferromagnetic Heusler alloy (Cu$_2$MnAl). Nutators and precession fields were used to control the direction of the incident and final neutron polarization. The scattered polarization was analyzed with a $^3$He neutron spin filter. The analysing efficiency of the polarization, which decays over time, was corrected from several measurements of $P_{zz}$ on the (0,0,2)  reflection. This reflection is purely nuclear and therefore does not affect the neutron spin. 

We first oriented the sample with ${\bf  b}$ perpendicular to the scattering plane, to access the $h0l$ reflections. The crystal was placed in a \qty{1}{\tesla} magnetic field parallel to $\textbf{b}$ to align the domain population of the magnetic structure, and then transferred to Cryopad. This step was essential, as it is not possible to distinguish the two possible magnetic structures without aligning the domain population \cite{pj1990determination}. We measured a series of reflections in this orientation in the commensurate phase at 295 K, then cooled the sample to \qty{160}{\kelvin} to measure in the IC phase. We next re-oriented the sample to have ${\bf c}$ perpendicular to the scattering plane to access the $hk0$ reflections, and remagnetized the sample at 295 K.

To investigate the influence of a magnetic field on the magnetic domains, we performed an unpolarized neutron diffraction experiment on the D23 instrument at the ILL \cite{Jacobsen2024}. The sample was mounted in a 15\,T vertical field magnet with $(h,0,l)$ as the scattering plane. 

We first heated the sample to 440 K, about 20 K above the Néel temperature, to reset the magnetic domains. We then measured along ($1,0,l$) in zero field in the high temperature commensurate phase and in the low temperature IC phase. We observed the phase transition by the appearance of satellite peaks, consistent with earlier reports \cite{Kouvel1964, Zimmer1972, Cable1993,Song2020,Chen2024,wang_flat_2023}. 

 We then applied a magnetic field of 1 T along $-b$ (labeled $-1$~T), and measured along $(1,0,l)$ for various temperatures upon heating, both in the commensurate and in the IC phase. We next heated the sample to the commensurate phase and applied a field of 10 T along $b$. If the domains were retained upon entering the IC phase, this should be observable as a change in the observed scattering pattern depending on the direction and magnitude of the applied magnetic field. 

DFT calculations were performed using the Atomic Simulation Environment (ASE) \cite{larsen2017atomic} in conjunction with the GPAW electronic structure code \cite{mortensen2005,enkovaara2010,mortensen2024gpaw}. The experimental lattice constants and atomic positions were taken from Ref. \cite{pj1990determination} to ensure realistic structural input and given the non-collinear antiferromagnetic order of \Sn{}, the local spin-density approximation (LSDA) \cite{von1972local} implemented within the locally-collinear approximation framework \cite{kubler1988local} was adopted for the
for the exchange-correlation potential. Self-consistent states were obtained using a $\kkk{}$-point density of at least 15 Å, a plane-wave cutoff of 800 eV, and the energy pr. unit cell for each iteration was converged below changes of 1 \textmu eV.

\section{Results}
\begin{figure}
    \includegraphics[width=1\linewidth]{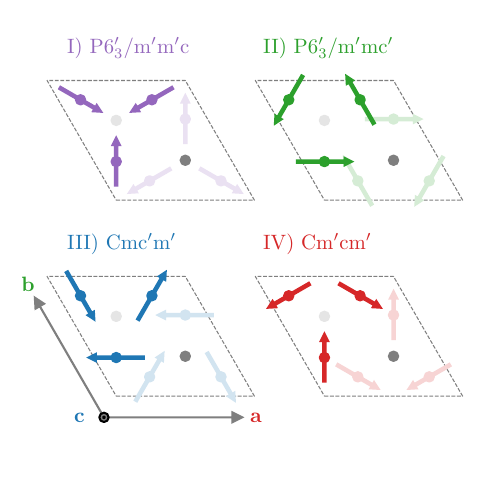}
    \caption{Four symmetry-allowed magnetic structures of \Sn{}. The Mn atoms are shown with colored circles, while Sn atoms are shown with gray. Faded atoms are at a lower layer than non-faded atoms.  We find that type III gives the best agreement with data. }
    \label{fig:Fig2}
\end{figure}

\begin{figure*}    
    \includegraphics[width=1\textwidth]{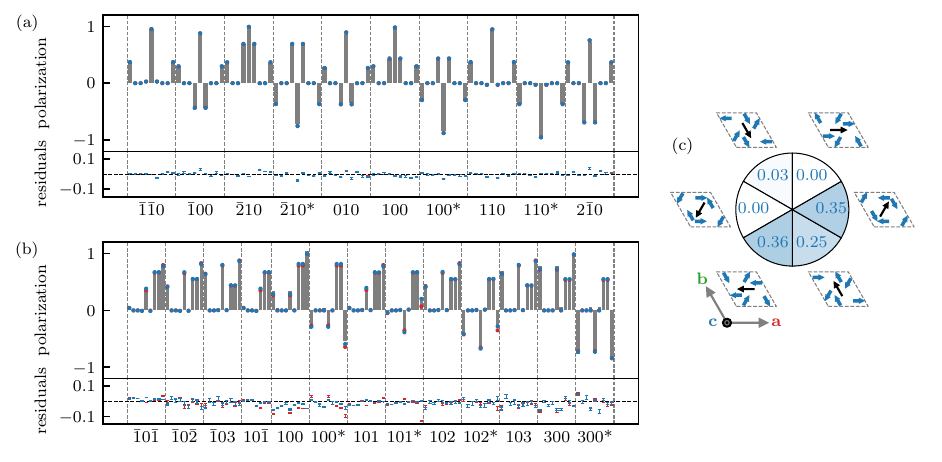}
    \caption{Comparison between the observed and calculated polarization matrix elements $P_{ij}$ for the Bragg peaks measured in the (a) $(hk0)$ and (b) $(h0l)$ scattering planes. For each reﬂection, the symbol (fit) and vertical bar (data) represent (from left to right) $P_{xx}$, $P_{yx}$, $P_{zx}$, $P_{xy}$, $P_{yy}$, $P_{zy}$, $P_{xz}$, $P_{yz}$, and $P_{zz}$. Reﬂections marked with an asterisk (*) are measurements that were repeated with the incident polarization reversed. The measured data are plotted as columns with errorbars, and the red and blue points represent the best fits for type III and IV structures, respectively. The difference between data and fit is shown at the bottom of the figures.
    (c) The domain distribution from our fit to (b). Each section in the pie chart shows the relative population of the domain, with the corresponding structure given outside the pie chart. The statistical uncertainty on these values is about 0.01. The black arrows show the direction of the small magnetic moment in the domain. The magnetic field was applied along ${\bf b}$. }
    \label{Fig:D3DataAndFit}
\end{figure*}

In Fig.~\ref{Fig:D3DataAndFit}(a) and (b) we depict the set of measured polarization matrix elements $P_{fi}$ in the two crystal orientations. For each peak, we show, from left to right, the measured $P_{xx}$, $P_{yx}$, $P_{zx}$, $P_{xy}$, $P_{yy}$, $P_{zy}$, $P_{xz}$, $P_{yz}$, and $P_{zz}$ as vertical bars. Colored points show the model.
The top panel contains reflections in the $(h,k,0)$ plane, while the bottom panel is for the $(h,0,l)$ plane. A few of the reflections, marked by an asterisk, were measured with reversed incident polarization. The fits were performed using MAG2POL \cite{qureshi_mag2pol_2019}, based on the Blume-Maleev equations \cite{blume_polarization_1963,maleev_scattering_1963}. The incident polarization of the monochromator was $P_i = 0.95$. Each crystal orientation was fitted separately to the four magnetic structures illustrated in Fig. \ref{fig:Fig2}. Fitting the two orientations simultaneously was not possible, as the sample was cooled to the IC phase between the different orientations, and the magnetic field was re-applied. As we discuss below, the domains are reset upon entering the IC phase. 
The only fit parameters were the population of the domains and the magnitude of the magnetic moments. 

\begin{table}[ht]
\centering
\begin{tabular}{llllll}
\hline 
\hline 
\multicolumn{2}{l}{  Magnetic structure }               & \textcolor{black}{I}      & \textcolor{black}{II}      & \textcolor{black}{III}      & \textcolor{black}{IV}       \\\hline
\multicolumn{1}{l}{\multirow{2}{*}{$\color{black}{(h,k,0)}$}} & $\chi_R^2$       & 4945   & 10368   & 16.00    & 16.02    \\ 
\multicolumn{1}{l}{}                         & ${ \mu}$ ($\mu_B$)  & 1.7(2) & 1.7(3)  & 2.447(9) & 2.442(5) \\
\multicolumn{1}{l}{\multirow{2}{*}{$\color{black}{(h,0,l)}$}} & $\chi_R^2$       & 18477  & 1701    & 37.14    & 73.03    \\ \multicolumn{1}{l}{}                         & ${\mu}$ ($\mu_B$) & 1.8(3) & 1.78(6) & 2.506(5) & 2.46(1)  \\ 
\multicolumn{2}{l}{ DFT Energy (meV)} & 7.416 & 8.382 & 0.000 & 0.000 \\ \hline \hline
\end{tabular}
    \caption{Results of fits of magnetic structure models for Mn$_3$Sn to SNP data. The reduced $\chi^2$ and ordered magnetic moment $\mu$ are listed for the two crystal orientations. The numbers in parentheses are standard deviations obtained from the fits. Type I and II have two domains each. Type III and IV have 6 domains each.
    Bottom row shows our DFT energies of the self-consistent state of the {\Sn} crystal (relative to the lowest energy).}
    \label{tab:Tab1}
\end{table}
 
Table~\ref{tab:Tab1} shows the reduced $\chi^2$ and magnetic moment $\mu$ for each of the four models for the two orientations. It is clear that only models III and IV give a reasonable fit to the data, and in the $(h,0,l)$ plane, type III is significantly better than type IV.

Our DFT calculations are in agreement with this result, as shown in the final row of Tab.~\ref{tab:Tab1}. The calculated energy of models I and II is significantly higher than that of models III and IV. Interestingly, the energy difference between type III and IV are below one $\mu$eV and the two structures can be regarded as degenerate within DFT. This is expected since sixth order magnetic anisotropy terms are required to distinguish the two states, and such higher order effects typically yield very small energy differences. 

To further confirm this, we also calculated the energy of the four models for \Ge{}, see Tab.~\ref{tab:Tab2}. We again find that type III and IV have the same energy.

\begin{table}[ht]
\centering
\begin{tabular}{lllll}
\hline 
\hline 
Magnetic structure & I & II & III & IV \\\hline 
\Sn{} & 7.416 & 8.382 & 0.000 & 0.000 \\
\Ge{} & 3.397 & 3.470 & 0.000 & 0.000 \\ \hline \hline
\end{tabular}
    \caption{
    Bottom row shows our DFT energies of the self-consistent state of the pristine {\Sn} and {\Ge} crystals (relative to the lowest energy for each crystal).}
    \label{tab:Tab2}    
\end{table}

For the best fit of type III, we find that three domains are approximately equally populated, while three have a negligible population. The domains and their distribution are illustrated in Fig.~\ref{Fig:D3DataAndFit}(c). red The exact values with uncertainties are given in Tab.~\ref{Table:Fitvalues} in the appendix.

We also measured selected peaks in the incommensurate phase at 160 K. Unfortunately, as discussed below, the data was not of sufficient quality to find a model that gave an unambiguous fit of that data.

Our experiment at D23, discussed below, confirms that the magnetic domains cannot be controlled by a magnetic field in the incommensurate phase. Therefore, all possible domains were equally populated, and most of the SNP terms canceled out.

To investigate the possibility of controlling the magnetic domains in the incommensurate phase, we measured the peaks from this phase as a function of temperature and magnetic field at D23. We extracted the position and amplitude of the peaks by fitting the peaks to Gaussians. 

\begin{figure}
    \centering
    \includegraphics[width=1\linewidth]{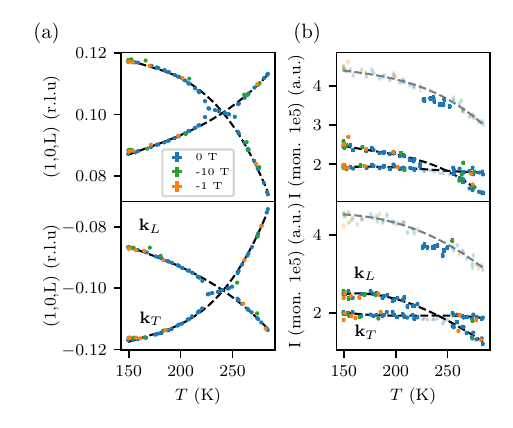}
    \caption{(a):  The position and (b) the intensity of the peaks in the incommensurate phase as a function of temperature. Blue, green and orange markers indicate data at zero field, 10 T and -1 T, respectively. Neither the peak position nor the integrated intensity is affected by the magnetic field. Lines are guides to the eye. 
    In a region around 230 K, the peaks are too close to be fitted separately, and a single peak was fitted instead. In (b), we show half the peak intensity in this region with faded markers. Outside that region, the faded markers indicate the sum of the two peaks.}
    \label{fig:Fig3}
\end{figure}

Fig.~\ref{fig:Fig3} shows a summary of the results of these fits. Fig.~\ref{fig:Fig3}(a) shows the peak position of the four incommensurate peaks at $(1,0,\pm \kkk_T)$ and $(1,0,\pm \kkk_L)$ 
as a function of temperature, with colors indicating the applied magnetic field. Fig.~\ref{fig:Fig3}(b) shows the intensity of the peaks.
Within statistical error, there is no difference between data taken in the three different field configurations. It is therefore clear that, although the magnetic domains can be controlled in the high temperature commensurate phase with a moderate magnetic field, the domains are not retained upon cooling into the incommensurate phase. The domains in the incommensurate phase can therefore not be controlled using any known methods, which unfortunately hinders the exact determination of the structure. 

\section{Discussion}
The clear preference for type III in our SNP analysis provides unambiguous evidence that the magnetic ground state of \Sn{} differs from that of isostructural \Ge{}, which adopts type IV~\cite{soh2020ground}. In that work, it was shown that sixth order anisotropy can distinguish between the models, and is allowed by symmetry. Our results imply that the  magnetic anisotropy has opposite sign in \Sn{}, as has indeed been proposed earlier \cite{Duan2015,Dasgupta2022,Chen2024}. In other words, the easy axis direction of \Ge{} is along the reciprocal lattice vectors ${\bf a}^*$, while for \Sn{} it is along the direct lattice vectors ${\bf a}$.

Our results are consistent with the work of Ref.~\cite{Duan2015}, which find a sixth-order anisotropy term of $K_6=-2210$ erg/cm$^3$, which corresponds to $2.9\times 10^{-5}$ meV/Mn. Such small energies cannot be distinguished by our DFT calculations, which indeed yield identical energies for type III and IV. 

Our experiments show that magnetic domains can be partially aligned using a moderate field in the high-temperature phase. Interestingly, three domains are selected with almost equal population, rather than one as might be expected. This can be understood as follows. Each of the six domains have a small net magnetic moment, indicated by arrows in Fig.~\ref{Fig:D3DataAndFit} (c)  \cite{Nagamiya1982}. This moment ($\boldsymbol\mu_\text{F}$) couples to the magnetic flux density (${\bf B}$) via a Zeeman term, $\mathcal{H} = -{\bf B} \cdot \boldsymbol\mu_\text{F}$. The net moment of three of the domains has a component parallel to the magnetic field, while the other three have an antiparallel component. The Zeeman energies of the domains with parallel moments are lower than those with antiparallel moments, and so the former three are preferentially populated.  Within this model we would expect the domain with moment along $\bbb$ to have the largest population, which we do not observe. It is unclear what causes this discrepancy. It may be related to the anisotropy not being uniform  \cite{Duan2015} or to small systematic errors in the data caused e.g. by slight depolarization of the beam at domain boundaries or due to the small ferromagnetic moment in the sample \cite{soh2020ground}.

In the previous SNP measurements of \Ge{}, type IV was the ground state and only two domains had significant populations of 60\% and 36\% \cite{soh2020ground}. There are two notable differences between the two experiments. First, as mentioned above, the easy axis direction of \Ge{} is along the reciprocal lattice vectors ${\bf a}^*$, while for \Sn{} it is along the direct lattice vectors ${\bf a}$. The magnetic field was applied along ${\bf b}$ in both experiments, and thus not along an easy axis for \Ge{}.
Second, the \Ge{} experiment was carried out at 2 K, whereas our experiment on \Sn{} had to be carried out at 295 K to be in the correct phase.  It is unclear whether the different domain populations are simply caused by temperature, or if other effects are at play. 

Either way, control of the magnetic domains plays a significant role in potential applications \cite{DalDin2024}. In thin films, it has furthermore been shown that the sign of the AHE (and therefore to some extent the magnetic domains) can be controlled electrically, which offers significant advantages for applications over control via magnetic field \cite{Tsai2020,Higo2022}.

Some research into spintronic applications use the chirality of spin spirals to store information \cite{Steinbrecher2018}. It is therefore of great interest to control the magnetic structure of \Sn{} in the low temperature phase. 

We attempted to determine the exact nature of the IC phase using SNP by measuring several of the IC peaks.  However, only the diagonal elements of the polarization matrix were non-zero, and we therefore did not have enough independent data points to obtain a unique solution. 
Although the available data were insufficient to fully resolve the structure, they are consistent with the model proposed in Ref.~\cite{wang_flat_2023}, reproduced in the appendix, Fig.~\ref{fig:Fig7}.  This model would allow for only two domains related by inversion along the $c$ axis. As we have shown here, these domains do not couple to a magnetic field, and we have therefore been unable to control them.

When moving along $c$ in the IC phase, the spins rotate by about 15 degrees between each layer (the exact rotation depends on the temperature). The magnetic domains are related by a rotation of 60 degrees, and the structure in the IC phase will therefore be similar to different domains in the commensurate phase at various layers (see Fig.~\ref{fig:Fig7} in the appendix).

We further note that models III and IV are related by a 30$^\circ$ rotation of all the spins.
 The magnetic structure will therefore also be similar to various domains of type IV in some layers.
This is consistent with the fact that type III and IV have very similar energies in the commensurate phase. 

In agreement with previous reports \cite{wang_flat_2023}, we find that the IC phase has two pairs of peaks associated with propagation vectors $\kkk_T\approx (0,0,\pm0.09)$ and ${\bf k} _L \approx (0,0, \pm0.11$), respectively. We also find a third peak, $k_3$ at $2k_L+k_T$ (see Fig.~\ref{fig:Fig6} in the appendix). This can be understood as a superposition of a longitudinal spin-density wave (${\bf k}_L$) and a coplanar helical magnetic order \cite{wang_flat_2023}. 

\section{Conclusion}
In conclusion, we have shown that the magnetic structure of \Sn{} in the high temperature phase is type III, i.e. space group Cmc'm' (see Fig.~\ref{fig:Fig2}). This is in contrast with Mn$_3$Ge which has the type IV structure (space group Cm'cm').  The ground state is most likely selected by sixth-order anisotropy, which therefore has the opposite sign in \Sn{} relative to \Ge{}. There are six possible magnetic domains in this phase, and a magnetic field applied along ${\bf b}$ selects not one but three of them with roughly equal population. Upon cooling below $\sim 290$ K, \Sn{} enters an incommensurate  phase, which is completely decoupled from external magnetic fields. This means that there is currently no known method of controlling the domain distribution in this phase. It would be interesting to pursue other routes than external magnetic fields to manipulate the domains in this phase, such as pulsed currents as demonstrated for MnAu$_2$ thin films \cite{Masuda2024}  or electric fields as demonstrated for the helical magnetic phase of CuO \cite{Babkevich2012}.

\begin{acknowledgments}
We thank Christian Dam Vedel for help with the experiments and for illuminating discussions. We thank Juan Manuel Parez Mato for pointing out that two of the magnetic space groups were mislabeled.  
HJ was funded by the Carlsberg Foundation Grant No. cf20-0497.
We thank the Institut Laue-Langevin (ILL), France for providing neutron beam time, DOI: 10.5291/ILL-DATA.5-54-393 and DOI: 10.5291/ILL-DATA.5-41-1246.
We thank the Danish Agency for Science, Technology, and Innovation for funding the instrument center DanScatt.
Y.F.G. acknowledges the support by the Double First-Class Initiative Fund of ShanghaiTech University and the open research fund of Beijing National Laboratory for Condensed Matter Physics (2023BNLCMPKF002). A.T.B. is grateful for support from the Oxford--ShanghaiTech collaboration project.
\end{acknowledgments}

\clearpage

\clearpage
\appendix
\section{Magnetic susceptibility and transport measurements}

\begin{figure}
    \includegraphics[width=0.4\textwidth]{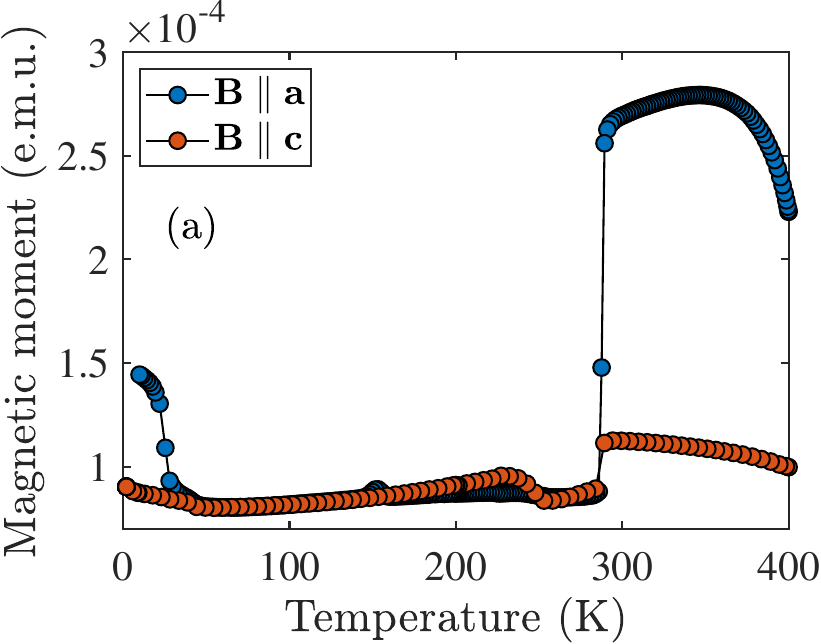}
    \includegraphics[width=0.4\textwidth]{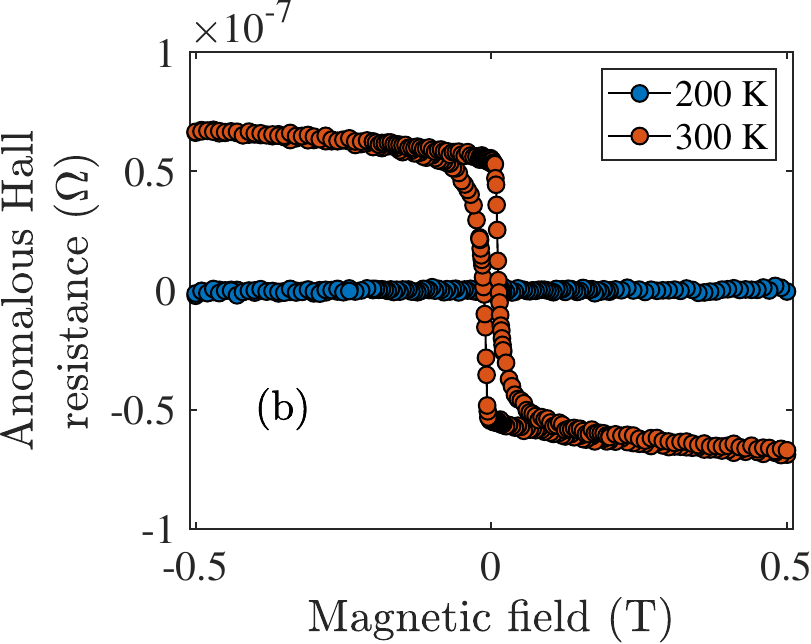}
    \caption{(a) Magnetic susceptibility of Mn$_3$Sn with the magnetic field of strength $\mu_0H=0.1$ T parallel to the $a$ (blue points) and $c$ (red points) axes. 
    (b) The Anomalous Hall resistance of \Sn{} at 300 K (red points) and at 200 K (blue points), showing the absence of AHE in the incommensurate phase.}
    \label{fig:Fig4}
\end{figure}

The magnetic susceptibility of the sample was measured at the \'{E}cole Polytechnique F\'{e}d\'{e}rale de Lausanne (EPFL), Switzerland, using a Superconducting Quantum Interference Device (SQUID) magnetometer from Quantum Design. The sample was cooled to 10\,K in zero field, then a field of 0.1\,T was applied and magnetic susceptibility  measurements were made while warming up to 400\,K. We measured with the magnetic field along the $a$ and $c$ axes. The results are shown in Fig.~\ref{fig:Fig4}. The magnetic phase transition at $\sim 290$ K is clearly visible. 

The Hall resistance was measured using the standard four-probe setup in a Physical Property Measurement System (PPMS) from Quantum Design, also at EPFL. Contacts were made on the surface of the sample using DuPont 4929N-100 silver paste from Delta Technologies. The Hall resistance was measured in magnetic fields along $a$ between -1 and 1\,T, at various temperatures between 10 and 400\,K, using a current of 8\,mA.
We estimated the background resistance (which includes the normal Hall contribution) by averaging over magnetic field and subtracting it, leaving only the anomalous Hall resistance.  

Due to the hexagonal shape of the sample, converting to resistivity would be prone to large systematic errors, and we therefore did not do this conversion. However, the numbers we find are close to previously published values \cite{nakatsuji2015large}. The results are shown in Fig.~\ref{fig:Fig4}.

\section{Domains from SNP fits}
Between re-orienting the sample from the $(h,0,l)$ plane to the $(h,k,0)$ plane in our SNP experiment at D3, we cooled the sample to measure the incommensurate phase. As shown in the main text, this reset the magnetic domains, and we therefore had to apply the magnetic field again to align the domains.

\begin{figure}
    \centering
    \includegraphics[width=1\linewidth]{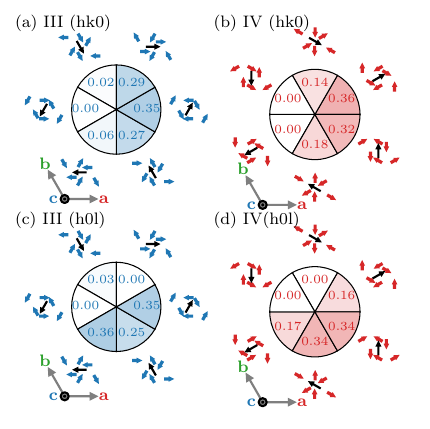}
    \caption{The domain distribution from our fits to models III and IV. Each subfigure shows the population of each domain in the pie chart and the spin directions of each domain. The black arrows show the direction of the small magnetic moment in the domain. 
    (a): Type III in the $(h,k,0)$ plane.
    (b) Type IV in the $(h,k,0)$ plane.
    (c) Type III in the $(h,0,l)$ plane.
    (d) Type IV in the $(h,0,l)$ plane.}
    \label{fig:Fig5}
\end{figure}

The fitted domain population are reported in \ref{Table:Fitvalues} and for the $(h,k,0)$ plane is shown in Fig.~\ref{fig:Fig5} (a). Here, we see the domains align along the (real space) [110] direction, which is equivalent to $b$. This is compatible with the magnetic field being applied along [110] in this measurement. 

\begin{table}
\centering
\caption{
Domain populations in Mn$_3$Sn for various fits. We use model III $(h0l)$ in the main text.}
\begin{tabular}{cllll}
\hline \hline
Domain & III $(hk0)$ & III $(h0l)$  & IV $(hk0)$ & IV $(h0l)$ \\
\hline
1 & 0.024(12)      & 0.027(12)    & 0.364(9)  & 0.159(11) \\
2 & 0.274(7)       & 0.255(7)     & 0.000      & 0.166(6)  \\
3 & 0.354(7)       & 0.353(4)     & 0.180(4)  & 0.341(1)  \\
4 & 0.000           & 0.0014(56)   & 0.136(3)  & 0.000      \\
5 & 0.062(5)       & 0.364(7)     & 0.000      & 0.000      \\
6 & 0.286(5)       & 0.000         & 0.320(7)  & 0.338(9)  \\
\hline \hline
\end{tabular}
\label{Table:Fitvalues}
\end{table}

Fig.~\ref{fig:Fig5} (b) and (d) we show the best fit results for type IV in the $(h,0,k)$ and $(h,0,l)$ plane, respectively. In type IV we consistently find four domains populated instead of three.

\section{Raw data from D23}
\begin{figure}
\includegraphics[]{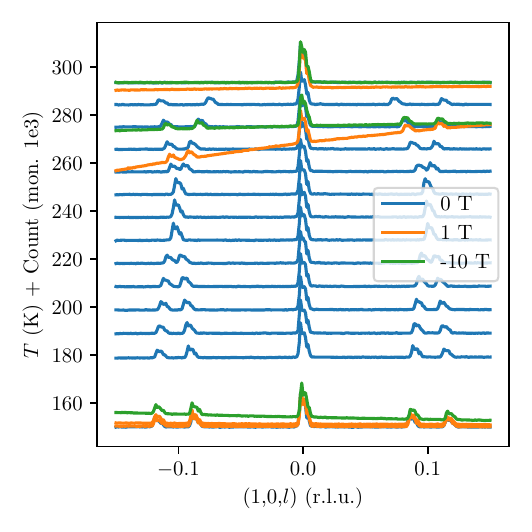}
\caption{Selected data from the D23 experiment with magnetic fields of: 0 T (blue), 1 T (orange) and $-10$ T (green). On the y-axis is a combination of the count and the temperature. Some measurements were made during heating and cooling. The two incommensurate peaks are visible near $l\sim \pm 0.1$, merging to one near 240 K. There is no evidence of an impact of the magnetic field. }
\label{Fig:D23_raw}
\end{figure}

In Fig.~\ref{Fig:D23_raw} we show the raw data from the D23 experiment, showing intensity as a function of $(1,0,l)$ and temperature.

\begin{figure}
    \centering
    \includegraphics[width=1\linewidth]{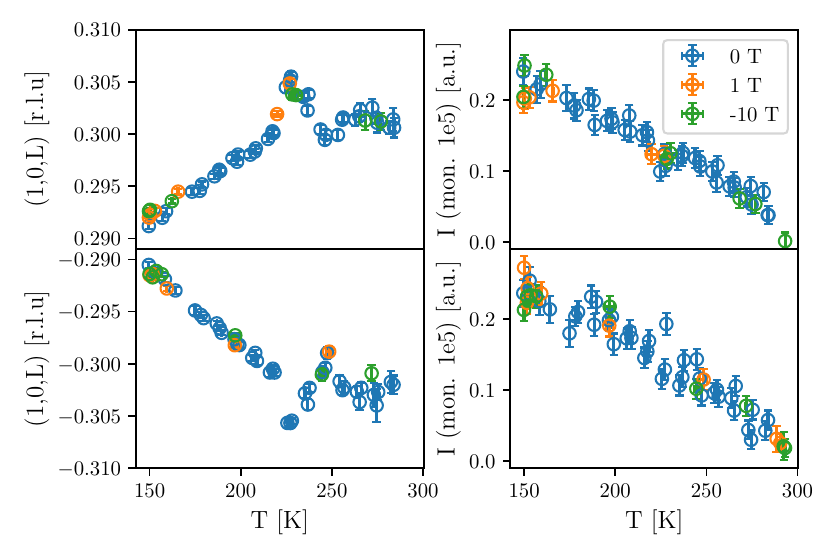}
    \caption{Left) Peak position of higher order peaks from D23 experiment with magnetic fields of: 0 T (blue), 1 T (orange) and $-10$ T (green). Right) Intensity of peaks as a function of temperature. Neither peak position nor intensity is affected by the magnetic field. }
    \label{fig:Fig6}
\end{figure}

In Fig.~\ref{fig:Fig6} we show the peak position and intensity of the higher order peaks. 

\section{Incommensurate structure}

\begin{figure}
    \centering
    \includegraphics[width=1\linewidth]{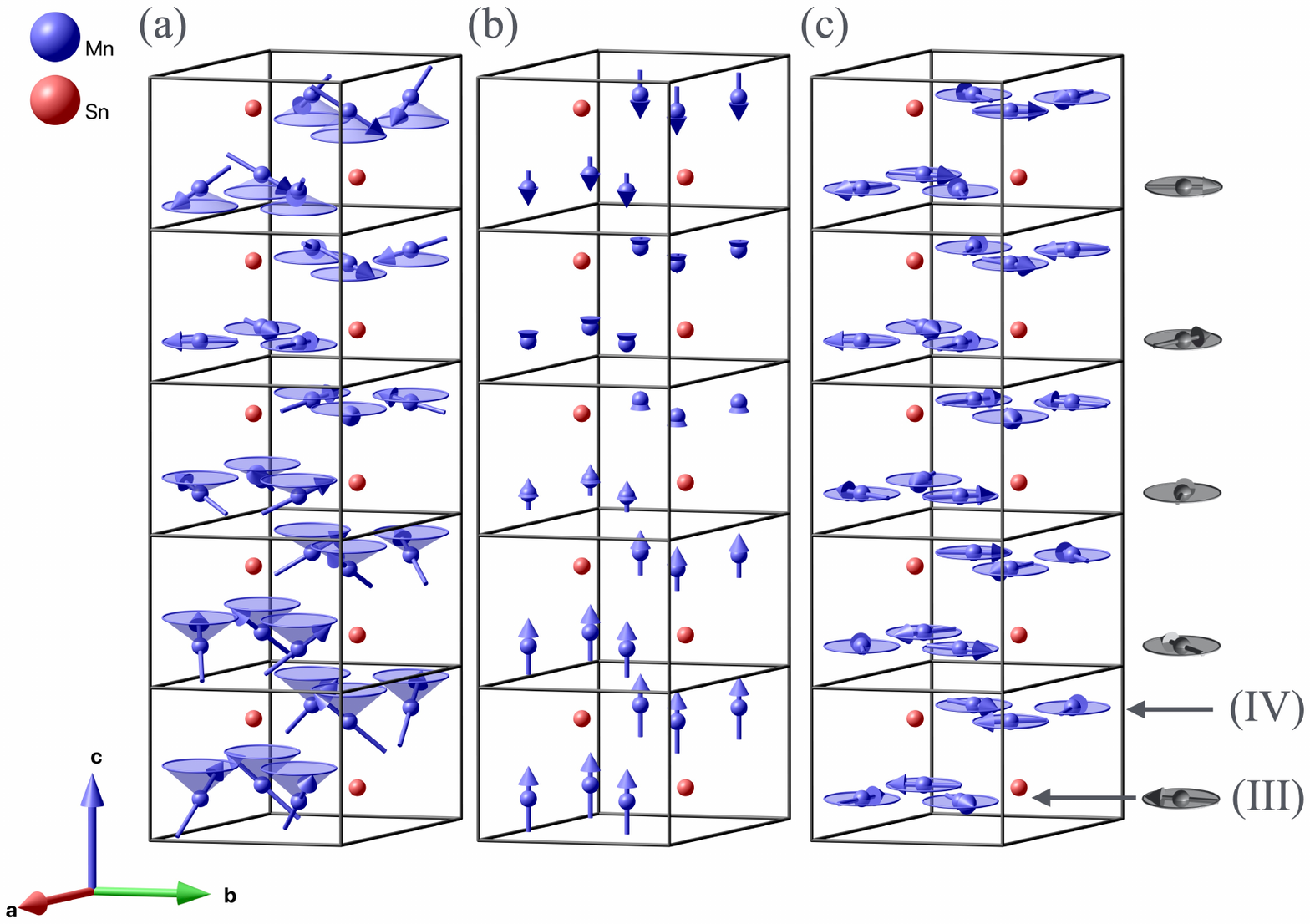}
    \caption{(a) Full incommensurate magnetic structure as a superposition of a spin density wave (b) and an in plane helical structure (c) adapted from Ref.~\cite{wang_flat_2023}. Due to the in plane rotation of the spins in the helical, and that structure (III) and (IV) are a $30^\circ$ rotation of each other, both spin configurations show up in the incommensurate structure, as indicated in (c). Spins in gray are the direction of the expected residual ferromagnetic moment for every second layer given only the in plane component (c).}
    \label{fig:Fig7}
\end{figure}


\begin{thebibliography}{47}%
	\makeatletter
	\providecommand \@ifxundefined [1]{%
		\@ifx{#1\undefined}
	}%
	\providecommand \@ifnum [1]{%
		\ifnum #1\expandafter \@firstoftwo
		\else \expandafter \@secondoftwo
		\fi
	}%
	\providecommand \@ifx [1]{%
		\ifx #1\expandafter \@firstoftwo
		\else \expandafter \@secondoftwo
		\fi
	}%
	\providecommand \natexlab [1]{#1}%
	\providecommand \enquote  [1]{``#1''}%
	\providecommand \bibnamefont  [1]{#1}%
	\providecommand \bibfnamefont [1]{#1}%
	\providecommand \citenamefont [1]{#1}%
	\providecommand \href@noop [0]{\@secondoftwo}%
	\providecommand \href [0]{\begingroup \@sanitize@url \@href}%
	\providecommand \@href[1]{\@@startlink{#1}\@@href}%
	\providecommand \@@href[1]{\endgroup#1\@@endlink}%
	\providecommand \@sanitize@url [0]{\catcode `\\12\catcode `\$12\catcode
		`\&12\catcode `\#12\catcode `\^12\catcode `\_12\catcode `\%12\relax}%
	\providecommand \@@startlink[1]{}%
	\providecommand \@@endlink[0]{}%
	\providecommand \url  [0]{\begingroup\@sanitize@url \@url }%
	\providecommand \@url [1]{\endgroup\@href {#1}{\urlprefix }}%
	\providecommand \urlprefix  [0]{URL }%
	\providecommand \Eprint [0]{\href }%
	\providecommand \doibase [0]{https://doi.org/}%
	\providecommand \selectlanguage [0]{\@gobble}%
	\providecommand \bibinfo  [0]{\@secondoftwo}%
	\providecommand \bibfield  [0]{\@secondoftwo}%
	\providecommand \translation [1]{[#1]}%
	\providecommand \BibitemOpen [0]{}%
	\providecommand \bibitemStop [0]{}%
	\providecommand \bibitemNoStop [0]{.\EOS\space}%
	\providecommand \EOS [0]{\spacefactor3000\relax}%
	\providecommand \BibitemShut  [1]{\csname bibitem#1\endcsname}%
	\let\auto@bib@innerbib\@empty
	%</preamble>
	\bibitem [{\citenamefont {Nakatsuji}\ \emph {et~al.}(2015)\citenamefont
		{Nakatsuji}, \citenamefont {Kiyohara},\ and\ \citenamefont
		{Higo}}]{nakatsuji2015large}%
	\BibitemOpen
	\bibfield  {author} {\bibinfo {author} {\bibfnamefont {S.}~\bibnamefont
			{Nakatsuji}}, \bibinfo {author} {\bibfnamefont {N.}~\bibnamefont
			{Kiyohara}},\ and\ \bibinfo {author} {\bibfnamefont {T.}~\bibnamefont
			{Higo}},\ }\bibfield  {title} {\bibinfo {title} {Large anomalous hall effect
			in a non-collinear antiferromagnet at room temperature},\ }\href@noop {}
	{\bibfield  {journal} {\bibinfo  {journal} {Nature}\ }\textbf {\bibinfo
			{volume} {527}},\ \bibinfo {pages} {212} (\bibinfo {year}
		{2015})}\BibitemShut {NoStop}%
	\bibitem [{\citenamefont {Kiyohara}\ \emph {et~al.}(2016)\citenamefont
		{Kiyohara}, \citenamefont {Tomita},\ and\ \citenamefont
		{Nakatsuji}}]{kiyohara2016giant}%
	\BibitemOpen
	\bibfield  {author} {\bibinfo {author} {\bibfnamefont {N.}~\bibnamefont
			{Kiyohara}}, \bibinfo {author} {\bibfnamefont {T.}~\bibnamefont {Tomita}},\
		and\ \bibinfo {author} {\bibfnamefont {S.}~\bibnamefont {Nakatsuji}},\
	}\bibfield  {title} {\bibinfo {title} {Giant anomalous {Hall} effect in the
			chiral antiferromagnet {Mn$_3$Ge}},\ }\href@noop {} {\bibfield  {journal}
		{\bibinfo  {journal} {Physical Review Applied}\ }\textbf {\bibinfo {volume}
			{5}},\ \bibinfo {pages} {064009} (\bibinfo {year} {2016})}\BibitemShut
	{NoStop}%
	\bibitem [{\citenamefont {Nayak}\ \emph {et~al.}(2016)\citenamefont {Nayak},
		\citenamefont {Fischer}, \citenamefont {Sun}, \citenamefont {Yan},
		\citenamefont {Karel}, \citenamefont {Komarek}, \citenamefont {Shekhar},
		\citenamefont {Kumar}, \citenamefont {Schnelle}, \citenamefont {K{\"u}bler}
		\emph {et~al.}}]{nayak2016large}%
	\BibitemOpen
	\bibfield  {author} {\bibinfo {author} {\bibfnamefont {A.~K.}\ \bibnamefont
			{Nayak}}, \bibinfo {author} {\bibfnamefont {J.~E.}\ \bibnamefont {Fischer}},
		\bibinfo {author} {\bibfnamefont {Y.}~\bibnamefont {Sun}}, \bibinfo {author}
		{\bibfnamefont {B.}~\bibnamefont {Yan}}, \bibinfo {author} {\bibfnamefont
			{J.}~\bibnamefont {Karel}}, \bibinfo {author} {\bibfnamefont {A.~C.}\
			\bibnamefont {Komarek}}, \bibinfo {author} {\bibfnamefont {C.}~\bibnamefont
			{Shekhar}}, \bibinfo {author} {\bibfnamefont {N.}~\bibnamefont {Kumar}},
		\bibinfo {author} {\bibfnamefont {W.}~\bibnamefont {Schnelle}}, \bibinfo
		{author} {\bibfnamefont {J.}~\bibnamefont {K{\"u}bler}}, \emph {et~al.},\
	}\bibfield  {title} {\bibinfo {title} {Large anomalous {Hall} effect driven
			by a nonvanishing {Berry} curvature in the noncollinear antiferromagnet
			{Mn$_3$Ge}},\ }\href@noop {} {\bibfield  {journal} {\bibinfo  {journal}
			{Science advances}\ }\textbf {\bibinfo {volume} {2}},\ \bibinfo {pages}
		{e1501870} (\bibinfo {year} {2016})}\BibitemShut {NoStop}%
	\bibitem [{\citenamefont {Sung}\ \emph {et~al.}(2018)\citenamefont {Sung},
		\citenamefont {Ronning}, \citenamefont {Thompson},\ and\ \citenamefont
		{Bauer}}]{Sung2018}%
	\BibitemOpen
	\bibfield  {author} {\bibinfo {author} {\bibfnamefont {N.~H.}\ \bibnamefont
			{Sung}}, \bibinfo {author} {\bibfnamefont {F.}~\bibnamefont {Ronning}},
		\bibinfo {author} {\bibfnamefont {J.~D.}\ \bibnamefont {Thompson}},\ and\
		\bibinfo {author} {\bibfnamefont {E.~D.}\ \bibnamefont {Bauer}},\ }\bibfield
	{title} {\bibinfo {title} {Magnetic phase dependence of the anomalous hall
			effect in {Mn$_3$Sn} single crystals},\ }\href
	{https://doi.org/10.1063/1.5021133} {\bibfield  {journal} {\bibinfo
			{journal} {Appl. Phys. Lett.}\ }\textbf {\bibinfo {volume} {112}},\ \bibinfo
		{pages} {132406} (\bibinfo {year} {2018})}\BibitemShut {NoStop}%
	\bibitem [{\citenamefont {Raju}\ \emph {et~al.}(2024)\citenamefont {Raju},
		\citenamefont {{Romero III}}, \citenamefont {Nishio-Hamane}, \citenamefont
		{Uesugi}, \citenamefont {Asakura}, \citenamefont {Tagay}, \citenamefont
		{Higo}, \citenamefont {Armitage}, \citenamefont {Broholm},\ and\
		\citenamefont {Nakatsuji}}]{Raju2024}%
	\BibitemOpen
	\bibfield  {author} {\bibinfo {author} {\bibfnamefont {M.}~\bibnamefont
			{Raju}}, \bibinfo {author} {\bibfnamefont {R.}~\bibnamefont {{Romero III}}},
		\bibinfo {author} {\bibfnamefont {D.}~\bibnamefont {Nishio-Hamane}}, \bibinfo
		{author} {\bibfnamefont {R.}~\bibnamefont {Uesugi}}, \bibinfo {author}
		{\bibfnamefont {M.}~\bibnamefont {Asakura}}, \bibinfo {author} {\bibfnamefont
			{Z.}~\bibnamefont {Tagay}}, \bibinfo {author} {\bibfnamefont
			{T.}~\bibnamefont {Higo}}, \bibinfo {author} {\bibfnamefont {N.~P.}\
			\bibnamefont {Armitage}}, \bibinfo {author} {\bibfnamefont {C.}~\bibnamefont
			{Broholm}},\ and\ \bibinfo {author} {\bibfnamefont {S.}~\bibnamefont
			{Nakatsuji}},\ }\bibfield  {title} {\bibinfo {title} {Anisotropic anomalous
			transport in the kagome-based topological antiferromagnetic {Mn$_3$Ga}
			epitaxial thin films},\ }\href
	{https://doi.org/10.1103/PhysRevMaterials.8.014204} {\bibfield  {journal}
		{\bibinfo  {journal} {Phys. Rev. Materials}\ }\textbf {\bibinfo {volume}
			{8}},\ \bibinfo {pages} {014204} (\bibinfo {year} {2024})}\BibitemShut
	{NoStop}%
	\bibitem [{\citenamefont {Chen}\ \emph {et~al.}(2021)\citenamefont {Chen},
		\citenamefont {Tomita}, \citenamefont {Minami}, \citenamefont {Fu},
		\citenamefont {Koretsune}, \citenamefont {Kitatani}, \citenamefont
		{Muhammad}, \citenamefont {Nishio-Hamane}, \citenamefont {Ishii},
		\citenamefont {Ishii}, \citenamefont {Arita},\ and\ \citenamefont
		{Nakatsuji}}]{Chen2021a}%
	\BibitemOpen
	\bibfield  {author} {\bibinfo {author} {\bibfnamefont {T.}~\bibnamefont
			{Chen}}, \bibinfo {author} {\bibfnamefont {T.}~\bibnamefont {Tomita}},
		\bibinfo {author} {\bibfnamefont {S.}~\bibnamefont {Minami}}, \bibinfo
		{author} {\bibfnamefont {M.}~\bibnamefont {Fu}}, \bibinfo {author}
		{\bibfnamefont {T.}~\bibnamefont {Koretsune}}, \bibinfo {author}
		{\bibfnamefont {M.}~\bibnamefont {Kitatani}}, \bibinfo {author}
		{\bibfnamefont {I.}~\bibnamefont {Muhammad}}, \bibinfo {author}
		{\bibfnamefont {D.}~\bibnamefont {Nishio-Hamane}}, \bibinfo {author}
		{\bibfnamefont {R.}~\bibnamefont {Ishii}}, \bibinfo {author} {\bibfnamefont
			{F.}~\bibnamefont {Ishii}}, \bibinfo {author} {\bibfnamefont
			{R.}~\bibnamefont {Arita}},\ and\ \bibinfo {author} {\bibfnamefont
			{S.}~\bibnamefont {Nakatsuji}},\ }\bibfield  {title} {\bibinfo {title}
		{{Anomalous transport due to Weyl fermions in the chiral antiferromagnets
				Mn$_3$X, X = Sn, Ge}},\ }\href {https://doi.org/10.1038/s41467-020-20838-1}
	{\bibfield  {journal} {\bibinfo  {journal} {Nat. Commun.}\ }\textbf {\bibinfo
			{volume} {12}},\ \bibinfo {pages} {572} (\bibinfo {year} {2021})}\BibitemShut
	{NoStop}%
	\bibitem [{\citenamefont {Dasgupta}(2022)}]{Dasgupta2022}%
	\BibitemOpen
	\bibfield  {author} {\bibinfo {author} {\bibfnamefont {S.}~\bibnamefont
			{Dasgupta}},\ }\bibfield  {title} {\bibinfo {title} {{Tuning the transport
				properties of ${\mathrm{Mn}}_{3}\mathrm{Ge}$ through the effect of strain on
				its magnetism}},\ }\href {https://doi.org/10.1103/PhysRevB.106.064431}
	{\bibfield  {journal} {\bibinfo  {journal} {Phys. Rev. B}\ }\textbf {\bibinfo
			{volume} {106}},\ \bibinfo {pages} {064431} (\bibinfo {year}
		{2022})}\BibitemShut {NoStop}%
	\bibitem [{\citenamefont {Ikhlas}\ \emph {et~al.}(2022)\citenamefont {Ikhlas},
		\citenamefont {Dasgupta}, \citenamefont {Theuss}, \citenamefont {Higo},
		\citenamefont {Kittaka}, \citenamefont {Ramshaw}, \citenamefont
		{Tchernyshyov}, \citenamefont {Hicks},\ and\ \citenamefont
		{Nakatsuji}}]{Ikhlas2022}%
	\BibitemOpen
	\bibfield  {author} {\bibinfo {author} {\bibfnamefont {M.}~\bibnamefont
			{Ikhlas}}, \bibinfo {author} {\bibfnamefont {S.}~\bibnamefont {Dasgupta}},
		\bibinfo {author} {\bibfnamefont {F.}~\bibnamefont {Theuss}}, \bibinfo
		{author} {\bibfnamefont {T.}~\bibnamefont {Higo}}, \bibinfo {author}
		{\bibfnamefont {S.}~\bibnamefont {Kittaka}}, \bibinfo {author} {\bibfnamefont
			{B.~J.}\ \bibnamefont {Ramshaw}}, \bibinfo {author} {\bibfnamefont
			{O.}~\bibnamefont {Tchernyshyov}}, \bibinfo {author} {\bibfnamefont {C.~W.}\
			\bibnamefont {Hicks}},\ and\ \bibinfo {author} {\bibfnamefont
			{S.}~\bibnamefont {Nakatsuji}},\ }\bibfield  {title} {\bibinfo {title}
		{Piezomagnetic switching of the anomalous hall effect in an antiferromagnet
			at room temperature},\ }\href {https://doi.org/10.1038/s41567-022-01645-5}
	{\bibfield  {journal} {\bibinfo  {journal} {Nature Physics}\ }\textbf
		{\bibinfo {volume} {18}},\ \bibinfo {pages} {1086} (\bibinfo {year}
		{2022})}\BibitemShut {NoStop}%
	\bibitem [{\citenamefont {Baltz}\ \emph {et~al.}(2018)\citenamefont {Baltz},
		\citenamefont {Manchon}, \citenamefont {Tsoi}, \citenamefont {Moriyama},
		\citenamefont {Ono},\ and\ \citenamefont {Tserkovnyak}}]{Baltz2018}%
	\BibitemOpen
	\bibfield  {author} {\bibinfo {author} {\bibfnamefont {V.}~\bibnamefont
			{Baltz}}, \bibinfo {author} {\bibfnamefont {A.}~\bibnamefont {Manchon}},
		\bibinfo {author} {\bibfnamefont {M.}~\bibnamefont {Tsoi}}, \bibinfo {author}
		{\bibfnamefont {T.}~\bibnamefont {Moriyama}}, \bibinfo {author}
		{\bibfnamefont {T.}~\bibnamefont {Ono}},\ and\ \bibinfo {author}
		{\bibfnamefont {Y.}~\bibnamefont {Tserkovnyak}},\ }\bibfield  {title}
	{\bibinfo {title} {Antiferromagnetic spintronics},\ }\href
	{https://doi.org/10.1103/RevModPhys.90.015005} {\bibfield  {journal}
		{\bibinfo  {journal} {Reviews of Modern Physics}\ }\textbf {\bibinfo {volume}
			{90}},\ \bibinfo {pages} {15005} (\bibinfo {year} {2018})}\BibitemShut
	{NoStop}%
	\bibitem [{\citenamefont {Din}\ \emph {et~al.}(2024)\citenamefont {Din},
		\citenamefont {Amin}, \citenamefont {Wadley},\ and\ \citenamefont
		{Edmonds}}]{DalDin2024}%
	\BibitemOpen
	\bibfield  {author} {\bibinfo {author} {\bibfnamefont {A.~D.}\ \bibnamefont
			{Din}}, \bibinfo {author} {\bibfnamefont {O.~J.}\ \bibnamefont {Amin}},
		\bibinfo {author} {\bibfnamefont {P.}~\bibnamefont {Wadley}},\ and\ \bibinfo
		{author} {\bibfnamefont {K.~W.}\ \bibnamefont {Edmonds}},\ }\bibfield
	{title} {\bibinfo {title} {Antiferromagnetic spintronics and beyond},\
	}\bibfield  {journal} {\bibinfo  {journal} {npj Spintronics}\ }\textbf
	{\bibinfo {volume} {2}},\ \href {https://doi.org/10.1038/s44306-024-00029-0}
	{10.1038/s44306-024-00029-0} (\bibinfo {year} {2024})\BibitemShut {NoStop}%
	\bibitem [{\citenamefont {Zheng}\ \emph {et~al.}(2024)\citenamefont {Zheng},
		\citenamefont {Zeng}, \citenamefont {Zhao}, \citenamefont {Shi},
		\citenamefont {Ren}, \citenamefont {Zhang}, \citenamefont {Jia},
		\citenamefont {Gu}, \citenamefont {Xiao}, \citenamefont {Zhou}, \citenamefont
		{Zhang}, \citenamefont {Lu}, \citenamefont {Wang}, \citenamefont {Zhao},
		\citenamefont {Li}, \citenamefont {Tay},\ and\ \citenamefont
		{Chen}}]{Zheng2024}%
	\BibitemOpen
	\bibfield  {author} {\bibinfo {author} {\bibfnamefont {Z.}~\bibnamefont
			{Zheng}}, \bibinfo {author} {\bibfnamefont {T.}~\bibnamefont {Zeng}},
		\bibinfo {author} {\bibfnamefont {T.}~\bibnamefont {Zhao}}, \bibinfo {author}
		{\bibfnamefont {S.}~\bibnamefont {Shi}}, \bibinfo {author} {\bibfnamefont
			{L.}~\bibnamefont {Ren}}, \bibinfo {author} {\bibfnamefont {T.}~\bibnamefont
			{Zhang}}, \bibinfo {author} {\bibfnamefont {L.}~\bibnamefont {Jia}}, \bibinfo
		{author} {\bibfnamefont {Y.}~\bibnamefont {Gu}}, \bibinfo {author}
		{\bibfnamefont {R.}~\bibnamefont {Xiao}}, \bibinfo {author} {\bibfnamefont
			{H.}~\bibnamefont {Zhou}}, \bibinfo {author} {\bibfnamefont {Q.}~\bibnamefont
			{Zhang}}, \bibinfo {author} {\bibfnamefont {J.}~\bibnamefont {Lu}}, \bibinfo
		{author} {\bibfnamefont {G.}~\bibnamefont {Wang}}, \bibinfo {author}
		{\bibfnamefont {C.}~\bibnamefont {Zhao}}, \bibinfo {author} {\bibfnamefont
			{H.}~\bibnamefont {Li}}, \bibinfo {author} {\bibfnamefont {B.~K.}\
			\bibnamefont {Tay}},\ and\ \bibinfo {author} {\bibfnamefont {J.}~\bibnamefont
			{Chen}},\ }\bibfield  {title} {\bibinfo {title} {Effective electrical
			manipulation of a topological antiferromagnet by orbital torques},\ }\href
	{https://doi.org/10.1038/s41467-024-45109-1} {\bibfield  {journal} {\bibinfo
			{journal} {Nat. Commun.}\ }\textbf {\bibinfo {volume} {15}},\ \bibinfo
		{pages} {745} (\bibinfo {year} {2024})}\BibitemShut {NoStop}%
	\bibitem [{\citenamefont {Chen}\ \emph {et~al.}(2014)\citenamefont {Chen},
		\citenamefont {Niu},\ and\ \citenamefont {MacDonald}}]{chen2014anomalous}%
	\BibitemOpen
	\bibfield  {author} {\bibinfo {author} {\bibfnamefont {H.}~\bibnamefont
			{Chen}}, \bibinfo {author} {\bibfnamefont {Q.}~\bibnamefont {Niu}},\ and\
		\bibinfo {author} {\bibfnamefont {A.~H.}\ \bibnamefont {MacDonald}},\
	}\bibfield  {title} {\bibinfo {title} {Anomalous {Hall} effect arising from
			noncollinear antiferromagnetism},\ }\href@noop {} {\bibfield  {journal}
		{\bibinfo  {journal} {Phys. Rev. Lett.}\ }\textbf {\bibinfo {volume} {112}},\
		\bibinfo {pages} {017205} (\bibinfo {year} {2014})}\BibitemShut {NoStop}%
	\bibitem [{\citenamefont {Li}\ \emph {et~al.}(2023)\citenamefont {Li},
		\citenamefont {Koo}, \citenamefont {Zhu}, \citenamefont {Behnia},\ and\
		\citenamefont {Yan}}]{Li2023}%
	\BibitemOpen
	\bibfield  {author} {\bibinfo {author} {\bibfnamefont {X.}~\bibnamefont
			{Li}}, \bibinfo {author} {\bibfnamefont {J.}~\bibnamefont {Koo}}, \bibinfo
		{author} {\bibfnamefont {Z.}~\bibnamefont {Zhu}}, \bibinfo {author}
		{\bibfnamefont {K.}~\bibnamefont {Behnia}},\ and\ \bibinfo {author}
		{\bibfnamefont {B.}~\bibnamefont {Yan}},\ }\bibfield  {title} {\bibinfo
		{title} {{Field-linear anomalous Hall effect and Berry curvature induced by
				spin chirality in the kagome antiferromagnet Mn$_3$Sn}},\ }\href
	{https://doi.org/10.1038/s41467-023-37076-w} {\bibfield  {journal} {\bibinfo
			{journal} {Nat. Commun.}\ }\textbf {\bibinfo {volume} {14}},\ \bibinfo
		{pages} {1642} (\bibinfo {year} {2023})}\BibitemShut {NoStop}%
	\bibitem [{\citenamefont {Boothroyd}(2022)}]{Boothroyd2022}%
	\BibitemOpen
	\bibfield  {author} {\bibinfo {author} {\bibfnamefont {A.~T.}\ \bibnamefont
			{Boothroyd}},\ }\bibfield  {title} {\bibinfo {title} {Topological electronic
			bands in crystalline solids},\ }\href
	{https://doi.org/10.1080/00107514.2023.2251764} {\bibfield  {journal}
		{\bibinfo  {journal} {Contemporary Physics}\ }\textbf {\bibinfo {volume}
			{63}},\ \bibinfo {pages} {305} (\bibinfo {year} {2022})}\BibitemShut
	{NoStop}%
	\bibitem [{\citenamefont {Ikhlas}\ \emph {et~al.}(2020)\citenamefont {Ikhlas},
		\citenamefont {Tomita},\ and\ \citenamefont {Nakatsuji}}]{Ikhlas2020}%
	\BibitemOpen
	\bibfield  {author} {\bibinfo {author} {\bibfnamefont {M.}~\bibnamefont
			{Ikhlas}}, \bibinfo {author} {\bibfnamefont {T.}~\bibnamefont {Tomita}},\
		and\ \bibinfo {author} {\bibfnamefont {S.}~\bibnamefont {Nakatsuji}},\
	}\bibfield  {title} {\bibinfo {title} {{Sample Quality Dependence of the
				Magnetic Properties in Non-Collinear Antiferromagnet Mn$_3$Sn}},\ }in\ \href
	{https://doi.org/10.7566/JPSCP.30.011177} {\emph {\bibinfo {booktitle}
			{Proceedings of the International Conference on Strongly Correlated Electron
				Systems (SCES2019)}}},\ Vol.\ \bibinfo {volume} {011177}\ (\bibinfo
	{publisher} {Journal of the Physical Society of Japan},\ \bibinfo {year}
	{2020})\ pp.\ \bibinfo {pages} {1--5}\BibitemShut {NoStop}%
	\bibitem [{\citenamefont {Park}\ \emph {et~al.}(2025)\citenamefont {Park},
		\citenamefont {Kim}, \citenamefont {Cho}, \citenamefont {Choi}, \citenamefont
		{Kwon}, \citenamefont {Seo},\ and\ \citenamefont {Park}}]{Park2025}%
	\BibitemOpen
	\bibfield  {author} {\bibinfo {author} {\bibfnamefont {J.}~\bibnamefont
			{Park}}, \bibinfo {author} {\bibfnamefont {W.~Y.}\ \bibnamefont {Kim}},
		\bibinfo {author} {\bibfnamefont {B.}~\bibnamefont {Cho}}, \bibinfo {author}
		{\bibfnamefont {W.~J.}\ \bibnamefont {Choi}}, \bibinfo {author}
		{\bibfnamefont {Y.~S.}\ \bibnamefont {Kwon}}, \bibinfo {author}
		{\bibfnamefont {J.}~\bibnamefont {Seo}},\ and\ \bibinfo {author}
		{\bibfnamefont {K.}~\bibnamefont {Park}},\ }\bibfield  {title} {\bibinfo
		{title} {{Nominal kagome antiferromagnetic Mn$_3$Sn: effects of excess Mn and
				its novel synthesis method}},\ }\href {https://doi.org/10.1039/d5tc00455a}
	{\bibfield  {journal} {\bibinfo  {journal} {J. Mater. Chem. C}\ }\textbf
		{\bibinfo {volume} {13}},\ \bibinfo {pages} {11869} (\bibinfo {year}
		{2025})}\BibitemShut {NoStop}%
	\bibitem [{\citenamefont {Kouvel}\ and\ \citenamefont
		{Kasper}(1964)}]{Kouvel1964}%
	\BibitemOpen
	\bibfield  {author} {\bibinfo {author} {\bibfnamefont {J.~S.}\ \bibnamefont
			{Kouvel}}\ and\ \bibinfo {author} {\bibfnamefont {J.~S.}\ \bibnamefont
			{Kasper}},\ }\bibfield  {title} {\bibinfo {title} {Triangular spin
			configuration in the antiferromagnetic intermetallic compounds {Mn$_3$Sn},
			{Mn$_3$Ge} and {Mn$_3$Rh}},\ }in\ \href@noop {} {\emph {\bibinfo {booktitle}
			{Proceedings of the International Conference on Magnetism}}}\ (\bibinfo
	{publisher} {Institute of Physics in association with Proceedings of the
		Physical Society},\ \bibinfo {address} {Nottingham, England},\ \bibinfo
	{year} {1964})\ pp.\ \bibinfo {pages} {169--170}\BibitemShut {NoStop}%
	\bibitem [{\citenamefont {Zimmer}\ and\ \citenamefont
		{Kr{\'e}n}(1972)}]{Zimmer1972}%
	\BibitemOpen
	\bibfield  {author} {\bibinfo {author} {\bibfnamefont {G.~J.}\ \bibnamefont
			{Zimmer}}\ and\ \bibinfo {author} {\bibfnamefont {E.}~\bibnamefont
			{Kr{\'e}n}},\ }\bibfield  {title} {\bibinfo {title} {Investigation of the
			magnetic phase transformation in mn$_3$sn},\ }in\ \href
	{https://doi.org/10.1063/1.3699489} {\emph {\bibinfo {booktitle} {AIP
				Conference Proceedings}}},\ Vol.~\bibinfo {volume} {5}\ (\bibinfo
	{publisher} {American Institute of Physics},\ \bibinfo {address} {New York,
		NY, USA},\ \bibinfo {year} {1972})\ pp.\ \bibinfo {pages}
	{513--516}\BibitemShut {NoStop}%
	\bibitem [{\citenamefont {Cable}\ \emph {et~al.}(1993)\citenamefont {Cable},
		\citenamefont {Wakabayashi},\ and\ \citenamefont {Radhakrishna}}]{Cable1993}%
	\BibitemOpen
	\bibfield  {author} {\bibinfo {author} {\bibfnamefont {J.~W.}\ \bibnamefont
			{Cable}}, \bibinfo {author} {\bibfnamefont {N.}~\bibnamefont {Wakabayashi}},\
		and\ \bibinfo {author} {\bibfnamefont {P.}~\bibnamefont {Radhakrishna}},\
	}\bibfield  {title} {\bibinfo {title} {{A neutron study of the magnetic
				structure of Mn$_3$Sn}},\ }\href
	{https://doi.org/10.1016/0038-1098(93)90400-H} {\bibfield  {journal}
		{\bibinfo  {journal} {Solid State Communications}\ }\textbf {\bibinfo
			{volume} {88}},\ \bibinfo {pages} {161} (\bibinfo {year} {1993})}\BibitemShut
	{NoStop}%
	\bibitem [{\citenamefont {Song}\ \emph {et~al.}(2020)\citenamefont {Song},
		\citenamefont {Hao}, \citenamefont {Wang}, \citenamefont {Zhang},
		\citenamefont {Huang}, \citenamefont {Xing},\ and\ \citenamefont
		{Chen}}]{Song2020}%
	\BibitemOpen
	\bibfield  {author} {\bibinfo {author} {\bibfnamefont {Y.}~\bibnamefont
			{Song}}, \bibinfo {author} {\bibfnamefont {Y.}~\bibnamefont {Hao}}, \bibinfo
		{author} {\bibfnamefont {S.}~\bibnamefont {Wang}}, \bibinfo {author}
		{\bibfnamefont {J.}~\bibnamefont {Zhang}}, \bibinfo {author} {\bibfnamefont
			{Q.}~\bibnamefont {Huang}}, \bibinfo {author} {\bibfnamefont
			{X.}~\bibnamefont {Xing}},\ and\ \bibinfo {author} {\bibfnamefont
			{J.}~\bibnamefont {Chen}},\ }\bibfield  {title} {\bibinfo {title}
		{{Complicated magnetic structure and its strong correlation with the
				anomalous Hall effect in Mn$_3$Sn}},\ }\href
	{https://doi.org/10.1103/PhysRevB.101.144422} {\bibfield  {journal} {\bibinfo
			{journal} {Phys. Rev. B}\ }\textbf {\bibinfo {volume} {101}},\ \bibinfo
		{pages} {144422} (\bibinfo {year} {2020})}\BibitemShut {NoStop}%
	\bibitem [{\citenamefont {Chen}\ \emph {et~al.}(2024)\citenamefont {Chen},
		\citenamefont {Gaudet}, \citenamefont {Marcus}, \citenamefont {Nomoto},
		\citenamefont {Chen}, \citenamefont {Tomita}, \citenamefont {Ikhlas},
		\citenamefont {Suzuki}, \citenamefont {Zhao}, \citenamefont {Chen},
		\citenamefont {Strempfer}, \citenamefont {Arita}, \citenamefont {Nakatsuji},\
		and\ \citenamefont {Broholm}}]{Chen2024}%
	\BibitemOpen
	\bibfield  {author} {\bibinfo {author} {\bibfnamefont {Y.}~\bibnamefont
			{Chen}}, \bibinfo {author} {\bibfnamefont {J.}~\bibnamefont {Gaudet}},
		\bibinfo {author} {\bibfnamefont {G.~G.}\ \bibnamefont {Marcus}}, \bibinfo
		{author} {\bibfnamefont {T.}~\bibnamefont {Nomoto}}, \bibinfo {author}
		{\bibfnamefont {T.}~\bibnamefont {Chen}}, \bibinfo {author} {\bibfnamefont
			{T.}~\bibnamefont {Tomita}}, \bibinfo {author} {\bibfnamefont
			{M.}~\bibnamefont {Ikhlas}}, \bibinfo {author} {\bibfnamefont {H.~S.}\
			\bibnamefont {Suzuki}}, \bibinfo {author} {\bibfnamefont {Y.}~\bibnamefont
			{Zhao}}, \bibinfo {author} {\bibfnamefont {W.~C.}\ \bibnamefont {Chen}},
		\bibinfo {author} {\bibfnamefont {J.}~\bibnamefont {Strempfer}}, \bibinfo
		{author} {\bibfnamefont {R.}~\bibnamefont {Arita}}, \bibinfo {author}
		{\bibfnamefont {S.}~\bibnamefont {Nakatsuji}},\ and\ \bibinfo {author}
		{\bibfnamefont {C.}~\bibnamefont {Broholm}},\ }\bibfield  {title} {\bibinfo
		{title} {Intertwined charge and spin density waves in a topological kagome
			material},\ }\href {https://doi.org/10.1103/PhysRevResearch.6.L032016}
	{\bibfield  {journal} {\bibinfo  {journal} {Phys. Rev. Res.}\ }\textbf
		{\bibinfo {volume} {6}},\ \bibinfo {pages} {L032016} (\bibinfo {year}
		{2024})}\BibitemShut {NoStop}%
	\bibitem [{\citenamefont {Wang}\ \emph {et~al.}(2023)\citenamefont {Wang},
		\citenamefont {Zhu}, \citenamefont {Yang}, \citenamefont {Meven},
		\citenamefont {Mi}, \citenamefont {Yi}, \citenamefont {Song}, \citenamefont
		{Mueller}, \citenamefont {Schmidt}, \citenamefont {Schmalzl}, \citenamefont
		{Ressouche}, \citenamefont {Xu}, \citenamefont {He}, \citenamefont {Shi},
		\citenamefont {Feng}, \citenamefont {Mokrousov}, \citenamefont {Blügel},
		\citenamefont {Roth},\ and\ \citenamefont {Su}}]{wang_flat_2023}%
	\BibitemOpen
	\bibfield  {author} {\bibinfo {author} {\bibfnamefont {X.}~\bibnamefont
			{Wang}}, \bibinfo {author} {\bibfnamefont {F.}~\bibnamefont {Zhu}}, \bibinfo
		{author} {\bibfnamefont {X.}~\bibnamefont {Yang}}, \bibinfo {author}
		{\bibfnamefont {M.}~\bibnamefont {Meven}}, \bibinfo {author} {\bibfnamefont
			{X.}~\bibnamefont {Mi}}, \bibinfo {author} {\bibfnamefont {C.}~\bibnamefont
			{Yi}}, \bibinfo {author} {\bibfnamefont {J.}~\bibnamefont {Song}}, \bibinfo
		{author} {\bibfnamefont {T.}~\bibnamefont {Mueller}}, \bibinfo {author}
		{\bibfnamefont {W.}~\bibnamefont {Schmidt}}, \bibinfo {author} {\bibfnamefont
			{K.}~\bibnamefont {Schmalzl}}, \bibinfo {author} {\bibfnamefont
			{E.}~\bibnamefont {Ressouche}}, \bibinfo {author} {\bibfnamefont
			{J.}~\bibnamefont {Xu}}, \bibinfo {author} {\bibfnamefont {M.}~\bibnamefont
			{He}}, \bibinfo {author} {\bibfnamefont {Y.}~\bibnamefont {Shi}}, \bibinfo
		{author} {\bibfnamefont {W.}~\bibnamefont {Feng}}, \bibinfo {author}
		{\bibfnamefont {Y.}~\bibnamefont {Mokrousov}}, \bibinfo {author}
		{\bibfnamefont {S.}~\bibnamefont {Blügel}}, \bibinfo {author} {\bibfnamefont
			{G.}~\bibnamefont {Roth}},\ and\ \bibinfo {author} {\bibfnamefont
			{Y.}~\bibnamefont {Su}},\ }\href {https://doi.org/10.48550/arXiv.2306.04312}
	{\bibinfo {title} {{Flat band-engineered spin-density wave and the emergent
				multi-k magnetic state in the topological kagome metal Mn$_3$Sn}}} (\bibinfo
	{year} {2023}),\ \bibinfo {note} {arXiv:2306.04312 [cond-mat]}\BibitemShut
	{NoStop}%
	\bibitem [{\citenamefont {Brown}\ \emph {et~al.}(1990)\citenamefont {Brown},
		\citenamefont {Nunez}, \citenamefont {Tasset}, \citenamefont {Forsyth},\ and\
		\citenamefont {Radhakrishna}}]{pj1990determination}%
	\BibitemOpen
	\bibfield  {author} {\bibinfo {author} {\bibfnamefont {P.~J.}\ \bibnamefont
			{Brown}}, \bibinfo {author} {\bibfnamefont {V.}~\bibnamefont {Nunez}},
		\bibinfo {author} {\bibfnamefont {F.}~\bibnamefont {Tasset}}, \bibinfo
		{author} {\bibfnamefont {J.~B.}\ \bibnamefont {Forsyth}},\ and\ \bibinfo
		{author} {\bibfnamefont {P.}~\bibnamefont {Radhakrishna}},\ }\bibfield
	{title} {\bibinfo {title} {Determination of the magnetic structure of
			{Mn$_3$Sn} using generalized neutron polarization analysis},\ }\href@noop {}
	{\bibfield  {journal} {\bibinfo  {journal} {Journal of Physics: Condensed
				Matter}\ }\textbf {\bibinfo {volume} {2}},\ \bibinfo {pages} {9409} (\bibinfo
		{year} {1990})}\BibitemShut {NoStop}%
	\bibitem [{\citenamefont {Tomiyoshi}\ \emph {et~al.}(1986)\citenamefont
		{Tomiyoshi}, \citenamefont {Abe}, \citenamefont {Yamaguchi}, \citenamefont
		{Yamauchi},\ and\ \citenamefont {Yamamoto}}]{Tomiyoshi1986}%
	\BibitemOpen
	\bibfield  {author} {\bibinfo {author} {\bibfnamefont {S.}~\bibnamefont
			{Tomiyoshi}}, \bibinfo {author} {\bibfnamefont {S.}~\bibnamefont {Abe}},
		\bibinfo {author} {\bibfnamefont {Y.}~\bibnamefont {Yamaguchi}}, \bibinfo
		{author} {\bibfnamefont {H.}~\bibnamefont {Yamauchi}},\ and\ \bibinfo
		{author} {\bibfnamefont {H.}~\bibnamefont {Yamamoto}},\ }\bibfield  {title}
	{\bibinfo {title} {{Triangular spin structure and weak ferromagnetism of
				Mn$_3$Sn at low temperature}},\ }\href
	{https://doi.org/10.1016/0304-8853(86)90353-7} {\bibfield  {journal}
		{\bibinfo  {journal} {Journal of Magnetism and Magnetic Materials}\ }\textbf
		{\bibinfo {volume} {54-57}},\ \bibinfo {pages} {1001} (\bibinfo {year}
		{1986})}\BibitemShut {NoStop}%
	\bibitem [{\citenamefont {Soh}\ \emph {et~al.}(2020)\citenamefont {Soh},
		\citenamefont {de~Juan}, \citenamefont {Qureshi}, \citenamefont {Jacobsen},
		\citenamefont {Wang}, \citenamefont {Guo},\ and\ \citenamefont
		{Boothroyd}}]{soh2020ground}%
	\BibitemOpen
	\bibfield  {author} {\bibinfo {author} {\bibfnamefont {J.-R.}\ \bibnamefont
			{Soh}}, \bibinfo {author} {\bibfnamefont {F.}~\bibnamefont {de~Juan}},
		\bibinfo {author} {\bibfnamefont {N.}~\bibnamefont {Qureshi}}, \bibinfo
		{author} {\bibfnamefont {H.}~\bibnamefont {Jacobsen}}, \bibinfo {author}
		{\bibfnamefont {H.-Y.}\ \bibnamefont {Wang}}, \bibinfo {author}
		{\bibfnamefont {Y.-F.}\ \bibnamefont {Guo}},\ and\ \bibinfo {author}
		{\bibfnamefont {A.~T.}\ \bibnamefont {Boothroyd}},\ }\bibfield  {title}
	{\bibinfo {title} {Ground-state magnetic structure of {Mn$_3$Ge}},\
	}\href@noop {} {\bibfield  {journal} {\bibinfo  {journal} {Physical Review
				B}\ }\textbf {\bibinfo {volume} {101}},\ \bibinfo {pages} {140411} (\bibinfo
		{year} {2020})}\BibitemShut {NoStop}%
	\bibitem [{Note1()}]{Note1}%
	\BibitemOpen
	\bibinfo {note} {Note that models (III) and (IV) were mislabelled as Pc'mm'
		and Pcm'm', respectively, in the original article by P.J Brown et al. \cite
		{pj1990determination}}\BibitemShut {NoStop}%
	\bibitem [{\citenamefont {Nagamiya}\ \emph {et~al.}(1982)\citenamefont
		{Nagamiya}, \citenamefont {Tomiyoshi},\ and\ \citenamefont
		{Yamaguchi}}]{Nagamiya1982}%
	\BibitemOpen
	\bibfield  {author} {\bibinfo {author} {\bibfnamefont {T.}~\bibnamefont
			{Nagamiya}}, \bibinfo {author} {\bibfnamefont {S.}~\bibnamefont
			{Tomiyoshi}},\ and\ \bibinfo {author} {\bibfnamefont {Y.}~\bibnamefont
			{Yamaguchi}},\ }\bibfield  {title} {\bibinfo {title} {Triangular spin
			configuration and weak ferromagnetism of mn$_3$sn and mn$_3$ge},\ }\href
	{https://doi.org/10.1016/0038-1098(82)90159-4} {\bibfield  {journal}
		{\bibinfo  {journal} {Solid State Communications}\ }\textbf {\bibinfo
			{volume} {42}},\ \bibinfo {pages} {385} (\bibinfo {year} {1982})}\BibitemShut
	{NoStop}%
	\bibitem [{\citenamefont {Higo}\ \emph {et~al.}(2022)\citenamefont {Higo},
		\citenamefont {Kondou}, \citenamefont {Nomoto}, \citenamefont {Shiga},
		\citenamefont {Sakamoto}, \citenamefont {Chen}, \citenamefont
		{Nishio-Hamane}, \citenamefont {Arita}, \citenamefont {Otani}, \citenamefont
		{Miwa},\ and\ \citenamefont {Nakatsuji}}]{Higo2022}%
	\BibitemOpen
	\bibfield  {author} {\bibinfo {author} {\bibfnamefont {T.}~\bibnamefont
			{Higo}}, \bibinfo {author} {\bibfnamefont {K.}~\bibnamefont {Kondou}},
		\bibinfo {author} {\bibfnamefont {T.}~\bibnamefont {Nomoto}}, \bibinfo
		{author} {\bibfnamefont {M.}~\bibnamefont {Shiga}}, \bibinfo {author}
		{\bibfnamefont {S.}~\bibnamefont {Sakamoto}}, \bibinfo {author}
		{\bibfnamefont {X.}~\bibnamefont {Chen}}, \bibinfo {author} {\bibfnamefont
			{D.}~\bibnamefont {Nishio-Hamane}}, \bibinfo {author} {\bibfnamefont
			{R.}~\bibnamefont {Arita}}, \bibinfo {author} {\bibfnamefont
			{Y.}~\bibnamefont {Otani}}, \bibinfo {author} {\bibfnamefont
			{S.}~\bibnamefont {Miwa}},\ and\ \bibinfo {author} {\bibfnamefont
			{S.}~\bibnamefont {Nakatsuji}},\ }\bibfield  {title} {\bibinfo {title}
		{Perpendicular full switching of chiral antiferromagnetic order by current},\
	}\href {https://doi.org/10.1038/s41586-022-04864-1} {\bibfield  {journal}
		{\bibinfo  {journal} {Nature}\ }\textbf {\bibinfo {volume} {607}},\ \bibinfo
		{pages} {474} (\bibinfo {year} {2022})}\BibitemShut {NoStop}%
	\bibitem [{\citenamefont {Jacobsen}\ \emph {et~al.}(2023)\citenamefont
		{Jacobsen}, \citenamefont {Cederholm}, \citenamefont {Krighaar},
		\citenamefont {Qureshi}, \citenamefont {Velamazan}, \citenamefont
		{Stunault},\ and\ \citenamefont {Vedel}}]{Jacobsen2023}%
	\BibitemOpen
	\bibfield  {author} {\bibinfo {author} {\bibfnamefont {H.}~\bibnamefont
			{Jacobsen}}, \bibinfo {author} {\bibfnamefont {J.}~\bibnamefont {Cederholm}},
		\bibinfo {author} {\bibfnamefont {K.}~\bibnamefont {Krighaar}}, \bibinfo
		{author} {\bibfnamefont {N.}~\bibnamefont {Qureshi}}, \bibinfo {author}
		{\bibfnamefont {J.~A.~R.}\ \bibnamefont {Velamazan}}, \bibinfo {author}
		{\bibfnamefont {A.}~\bibnamefont {Stunault}},\ and\ \bibinfo {author}
		{\bibfnamefont {C.}~\bibnamefont {Vedel}},\ }\href
	{https://doi.org/10.5291/ILL-DATA.5-54-393} {\bibinfo {title} {{Is the
				incommensurate magnetic order in Mn\textsubscript{3}Sn cycloidal or
				cosine?}}} (\bibinfo {year} {2023}),\ \bibinfo {note} {{DOI}:
		10.5291/ILL-DATA.5-54-393}\BibitemShut {NoStop}%
	\bibitem [{\citenamefont {Boothroyd}(2020)}]{Andrews_book}%
	\BibitemOpen
	\bibfield  {author} {\bibinfo {author} {\bibfnamefont {A.~T.}\ \bibnamefont
			{Boothroyd}},\ }\href@noop {} {\emph {\bibinfo {title} {Principles of Neutron
				Scattering from Condensed Matter}}}\ (\bibinfo  {publisher} {Oxford
		University Press, Oxford, UK},\ \bibinfo {year} {2020})\BibitemShut {NoStop}%
	\bibitem [{\citenamefont {Tasset}\ \emph {et~al.}(1999)\citenamefont {Tasset},
		\citenamefont {Brown}, \citenamefont {Leli\'{e}vre-Berna}, \citenamefont
		{Roberts}, \citenamefont {Pujol}, \citenamefont {Allibon},\ and\
		\citenamefont {Bourgeat-Lami}}]{Tasset1999}%
	\BibitemOpen
	\bibfield  {author} {\bibinfo {author} {\bibfnamefont {F.}~\bibnamefont
			{Tasset}}, \bibinfo {author} {\bibfnamefont {P.}~\bibnamefont {Brown}},
		\bibinfo {author} {\bibfnamefont {E.}~\bibnamefont {Leli\'{e}vre-Berna}},
		\bibinfo {author} {\bibfnamefont {T.}~\bibnamefont {Roberts}}, \bibinfo
		{author} {\bibfnamefont {S.}~\bibnamefont {Pujol}}, \bibinfo {author}
		{\bibfnamefont {J.}~\bibnamefont {Allibon}},\ and\ \bibinfo {author}
		{\bibfnamefont {E.}~\bibnamefont {Bourgeat-Lami}},\ }\bibfield  {title}
	{\bibinfo {title} {Spherical neutron polarimetry with {Cryopad-II}},\ }\href
	{https://doi.org/10.1016/S0921-4526(99)00029-0} {\bibfield  {journal}
		{\bibinfo  {journal} {Physica B}\ }\textbf {\bibinfo {volume} {267--268}},\
		\bibinfo {pages} {69} (\bibinfo {year} {1999})}\BibitemShut {NoStop}%
	\bibitem [{\citenamefont {Leli\'{e}vre-Berna}\ \emph
		{et~al.}(2005)\citenamefont {Leli\'{e}vre-Berna}, \citenamefont
		{Bourgeat-Lami}, \citenamefont {Fouilloux}, \citenamefont {Geffray},
		\citenamefont {Gibert}, \citenamefont {Kakurai}, \citenamefont {Kernavanois},
		\citenamefont {Longuet}, \citenamefont {Mantegezza}, \citenamefont
		{Nakamura}, \citenamefont {Pujol}, \citenamefont {Regnault}, \citenamefont
		{Tasset}, \citenamefont {Takeda}, \citenamefont {Thomas},\ and\ \citenamefont
		{Tonon}}]{Lelievre2005}%
	\BibitemOpen
	\bibfield  {author} {\bibinfo {author} {\bibfnamefont {E.}~\bibnamefont
			{Leli\'{e}vre-Berna}}, \bibinfo {author} {\bibfnamefont {E.}~\bibnamefont
			{Bourgeat-Lami}}, \bibinfo {author} {\bibfnamefont {P.}~\bibnamefont
			{Fouilloux}}, \bibinfo {author} {\bibfnamefont {B.}~\bibnamefont {Geffray}},
		\bibinfo {author} {\bibfnamefont {Y.}~\bibnamefont {Gibert}}, \bibinfo
		{author} {\bibfnamefont {K.}~\bibnamefont {Kakurai}}, \bibinfo {author}
		{\bibfnamefont {N.}~\bibnamefont {Kernavanois}}, \bibinfo {author}
		{\bibfnamefont {B.}~\bibnamefont {Longuet}}, \bibinfo {author} {\bibfnamefont
			{F.}~\bibnamefont {Mantegezza}}, \bibinfo {author} {\bibfnamefont
			{M.}~\bibnamefont {Nakamura}}, \bibinfo {author} {\bibfnamefont
			{S.}~\bibnamefont {Pujol}}, \bibinfo {author} {\bibfnamefont {L.-P.}\
			\bibnamefont {Regnault}}, \bibinfo {author} {\bibfnamefont {F.}~\bibnamefont
			{Tasset}}, \bibinfo {author} {\bibfnamefont {M.}~\bibnamefont {Takeda}},
		\bibinfo {author} {\bibfnamefont {M.}~\bibnamefont {Thomas}},\ and\ \bibinfo
		{author} {\bibfnamefont {X.}~\bibnamefont {Tonon}},\ }\bibfield  {title}
	{\bibinfo {title} {Advances in spherical neutron polarimetry with
			{Cryopad}},\ }\href {https://doi.org/10.1016/j.physb.2004.10.063} {\bibfield
		{journal} {\bibinfo  {journal} {Physica B}\ }\textbf {\bibinfo {volume}
			{356}},\ \bibinfo {pages} {131} (\bibinfo {year} {2005})}\BibitemShut
	{NoStop}%
	\bibitem [{\citenamefont {Jacobsen}\ \emph {et~al.}(2024)\citenamefont
		{Jacobsen}, \citenamefont {Cederholm}, \citenamefont {Qureshi}, \citenamefont
		{Ressouche}, \citenamefont {Velamazan},\ and\ \citenamefont
		{Vedel}}]{Jacobsen2024}%
	\BibitemOpen
	\bibfield  {author} {\bibinfo {author} {\bibfnamefont {H.}~\bibnamefont
			{Jacobsen}}, \bibinfo {author} {\bibfnamefont {J.}~\bibnamefont {Cederholm}},
		\bibinfo {author} {\bibfnamefont {N.}~\bibnamefont {Qureshi}}, \bibinfo
		{author} {\bibfnamefont {E.}~\bibnamefont {Ressouche}}, \bibinfo {author}
		{\bibfnamefont {J.~A.~R.}\ \bibnamefont {Velamazan}},\ and\ \bibinfo {author}
		{\bibfnamefont {C.}~\bibnamefont {Vedel}},\ }\href
	{https://doi.org/10.5291/ILL-DATA.5-41-1246} {\bibinfo {title} {{Controlling
				the magnetic domains of Mn\textsubscript{3}Sn with a magnetic field}}}
	(\bibinfo {year} {2024})\BibitemShut {NoStop}%
	\bibitem [{\citenamefont {Larsen}\ \emph {et~al.}(2017)\citenamefont {Larsen},
		\citenamefont {Mortensen}, \citenamefont {Blomqvist}, \citenamefont
		{Castelli}, \citenamefont {Christensen}, \citenamefont {Du{\l}ak},
		\citenamefont {Friis}, \citenamefont {Groves}, \citenamefont {Hammer},
		\citenamefont {Hargus} \emph {et~al.}}]{larsen2017atomic}%
	\BibitemOpen
	\bibfield  {author} {\bibinfo {author} {\bibfnamefont {A.~H.}\ \bibnamefont
			{Larsen}}, \bibinfo {author} {\bibfnamefont {J.~J.}\ \bibnamefont
			{Mortensen}}, \bibinfo {author} {\bibfnamefont {J.}~\bibnamefont
			{Blomqvist}}, \bibinfo {author} {\bibfnamefont {I.~E.}\ \bibnamefont
			{Castelli}}, \bibinfo {author} {\bibfnamefont {R.}~\bibnamefont
			{Christensen}}, \bibinfo {author} {\bibfnamefont {M.}~\bibnamefont
			{Du{\l}ak}}, \bibinfo {author} {\bibfnamefont {J.}~\bibnamefont {Friis}},
		\bibinfo {author} {\bibfnamefont {M.~N.}\ \bibnamefont {Groves}}, \bibinfo
		{author} {\bibfnamefont {B.}~\bibnamefont {Hammer}}, \bibinfo {author}
		{\bibfnamefont {C.}~\bibnamefont {Hargus}}, \emph {et~al.},\ }\bibfield
	{title} {\bibinfo {title} {{The atomic simulation environment—a Python
				library for working with atoms}},\ }\href@noop {} {\bibfield  {journal}
		{\bibinfo  {journal} {Journal of Physics: Condensed Matter}\ }\textbf
		{\bibinfo {volume} {29}},\ \bibinfo {pages} {273002} (\bibinfo {year}
		{2017})}\BibitemShut {NoStop}%
	\bibitem [{\citenamefont {Mortensen}\ \emph {et~al.}(2005)\citenamefont
		{Mortensen}, \citenamefont {Hansen},\ and\ \citenamefont
		{Jacobsen}}]{mortensen2005}%
	\BibitemOpen
	\bibfield  {author} {\bibinfo {author} {\bibfnamefont {J.~J.}\ \bibnamefont
			{Mortensen}}, \bibinfo {author} {\bibfnamefont {L.~B.}\ \bibnamefont
			{Hansen}},\ and\ \bibinfo {author} {\bibfnamefont {K.~W.}\ \bibnamefont
			{Jacobsen}},\ }\bibfield  {title} {\bibinfo {title} {Real-space grid
			implementation of the projector augmented wave method},\ }\href@noop {}
	{\bibfield  {journal} {\bibinfo  {journal} {Phys. Rev. B}\ }\textbf {\bibinfo
			{volume} {71}},\ \bibinfo {pages} {035109} (\bibinfo {year}
		{2005})}\BibitemShut {NoStop}%
	\bibitem [{\citenamefont {Enkovaara}\ \emph {et~al.}(2010)\citenamefont
		{Enkovaara}, \citenamefont {Rostgaard}, \citenamefont {Mortensen},
		\citenamefont {Chen}, \citenamefont {Du{\l}ak}, \citenamefont {Ferrighi},
		\citenamefont {Gavnholt}, \citenamefont {Glinsvad}, \citenamefont {Haikola},
		\citenamefont {Hansen} \emph {et~al.}}]{enkovaara2010}%
	\BibitemOpen
	\bibfield  {author} {\bibinfo {author} {\bibfnamefont {J.}~\bibnamefont
			{Enkovaara}}, \bibinfo {author} {\bibfnamefont {C.}~\bibnamefont
			{Rostgaard}}, \bibinfo {author} {\bibfnamefont {J.~J.}\ \bibnamefont
			{Mortensen}}, \bibinfo {author} {\bibfnamefont {J.}~\bibnamefont {Chen}},
		\bibinfo {author} {\bibfnamefont {M.}~\bibnamefont {Du{\l}ak}}, \bibinfo
		{author} {\bibfnamefont {L.}~\bibnamefont {Ferrighi}}, \bibinfo {author}
		{\bibfnamefont {J.}~\bibnamefont {Gavnholt}}, \bibinfo {author}
		{\bibfnamefont {C.}~\bibnamefont {Glinsvad}}, \bibinfo {author}
		{\bibfnamefont {V.}~\bibnamefont {Haikola}}, \bibinfo {author} {\bibfnamefont
			{H.~A.}\ \bibnamefont {Hansen}}, \emph {et~al.},\ }\bibfield  {title}
	{\bibinfo {title} {Electronic structure calculations with {GPAW}: a
			real-space implementation of theprojector augmented-wave method},\
	}\href@noop {} {\bibfield  {journal} {\bibinfo  {journal} {J. Phys.: Condens.
				Matter}\ }\textbf {\bibinfo {volume} {22}},\ \bibinfo {pages} {253202}
		(\bibinfo {year} {2010})}\BibitemShut {NoStop}%
	\bibitem [{\citenamefont {Mortensen}\ \emph {et~al.}(2024)\citenamefont
		{Mortensen}, \citenamefont {Larsen}, \citenamefont {Kuisma}, \citenamefont
		{Ivanov}, \citenamefont {Taghizadeh}, \citenamefont {Peterson}, \citenamefont
		{Haldar}, \citenamefont {Dohn}, \citenamefont {Sch{\"a}fer}, \citenamefont
		{J{\'o}nsson} \emph {et~al.}}]{mortensen2024gpaw}%
	\BibitemOpen
	\bibfield  {author} {\bibinfo {author} {\bibfnamefont {J.~J.}\ \bibnamefont
			{Mortensen}}, \bibinfo {author} {\bibfnamefont {A.~H.}\ \bibnamefont
			{Larsen}}, \bibinfo {author} {\bibfnamefont {M.}~\bibnamefont {Kuisma}},
		\bibinfo {author} {\bibfnamefont {A.~V.}\ \bibnamefont {Ivanov}}, \bibinfo
		{author} {\bibfnamefont {A.}~\bibnamefont {Taghizadeh}}, \bibinfo {author}
		{\bibfnamefont {A.}~\bibnamefont {Peterson}}, \bibinfo {author}
		{\bibfnamefont {A.}~\bibnamefont {Haldar}}, \bibinfo {author} {\bibfnamefont
			{A.~O.}\ \bibnamefont {Dohn}}, \bibinfo {author} {\bibfnamefont
			{C.}~\bibnamefont {Sch{\"a}fer}}, \bibinfo {author} {\bibfnamefont
			{E.~{\"O}.}\ \bibnamefont {J{\'o}nsson}}, \emph {et~al.},\ }\bibfield
	{title} {\bibinfo {title} {Gpaw: An open {Python} package for electronic
			structure calculations},\ }\href@noop {} {\bibfield  {journal} {\bibinfo
			{journal} {The Journal of Chemical Physics}\ }\textbf {\bibinfo {volume}
			{160}} (\bibinfo {year} {2024})}\BibitemShut {NoStop}%
	\bibitem [{\citenamefont {von Barth}\ and\ \citenamefont
		{Hedin}(1972)}]{von1972local}%
	\BibitemOpen
	\bibfield  {author} {\bibinfo {author} {\bibfnamefont {U.}~\bibnamefont {von
				Barth}}\ and\ \bibinfo {author} {\bibfnamefont {L.}~\bibnamefont {Hedin}},\
	}\bibfield  {title} {\bibinfo {title} {A local exchange-correlation potential
			for the spin polarized case. {I}},\ }\href@noop {} {\bibfield  {journal}
		{\bibinfo  {journal} {Journal of Physics C: Solid State Physics}\ }\textbf
		{\bibinfo {volume} {5}},\ \bibinfo {pages} {1629} (\bibinfo {year}
		{1972})}\BibitemShut {NoStop}%
	\bibitem [{\citenamefont {K{\"u}bler}\ \emph {et~al.}(1988)\citenamefont
		{K{\"u}bler}, \citenamefont {H{\"o}ck}, \citenamefont {Sticht},\ and\
		\citenamefont {Williams}}]{kubler1988local}%
	\BibitemOpen
	\bibfield  {author} {\bibinfo {author} {\bibfnamefont {J.}~\bibnamefont
			{K{\"u}bler}}, \bibinfo {author} {\bibfnamefont {K.-H.}\ \bibnamefont
			{H{\"o}ck}}, \bibinfo {author} {\bibfnamefont {J.}~\bibnamefont {Sticht}},\
		and\ \bibinfo {author} {\bibfnamefont {A.~R.}\ \bibnamefont {Williams}},\
	}\bibfield  {title} {\bibinfo {title} {Local spin-density functional theory
			of noncollinear magnetism},\ }\href@noop {} {\bibfield  {journal} {\bibinfo
			{journal} {Journal of applied physics}\ }\textbf {\bibinfo {volume} {63}},\
		\bibinfo {pages} {3482} (\bibinfo {year} {1988})}\BibitemShut {NoStop}%
	\bibitem [{\citenamefont {Qureshi}(2019)}]{qureshi_mag2pol_2019}%
	\BibitemOpen
	\bibfield  {author} {\bibinfo {author} {\bibfnamefont {N.}~\bibnamefont
			{Qureshi}},\ }\bibfield  {title} {\bibinfo {title} {{Mag2Pol}: a program for
			the analysis of spherical neutron polarimetry, flipping ratio and integrated
			intensity data},\ }\href {https://doi.org/10.1107/S1600576718016084}
	{\bibfield  {journal} {\bibinfo  {journal} {J. Appl. Crystallogr.}\ }\textbf
		{\bibinfo {volume} {52}},\ \bibinfo {pages} {175} (\bibinfo {year}
		{2019})}\BibitemShut {NoStop}%
	\bibitem [{\citenamefont {Blume}(1963)}]{blume_polarization_1963}%
	\BibitemOpen
	\bibfield  {author} {\bibinfo {author} {\bibfnamefont {M.}~\bibnamefont
			{Blume}},\ }\bibfield  {title} {\bibinfo {title} {Polarization {Effects} in
			the {Magnetic} {Elastic} {Scattering} of {Slow} {Neutrons}},\ }\href
	{https://doi.org/10.1103/PhysRev.130.1670} {\bibfield  {journal} {\bibinfo
			{journal} {Physical Review}\ }\textbf {\bibinfo {volume} {130}},\ \bibinfo
		{pages} {1670} (\bibinfo {year} {1963})}\BibitemShut {NoStop}%
	\bibitem [{\citenamefont {Maleev}\ \emph {et~al.}(1963)\citenamefont {Maleev},
		\citenamefont {Bar'yakhtar},\ and\ \citenamefont
		{Suris}}]{maleev_scattering_1963}%
	\BibitemOpen
	\bibfield  {author} {\bibinfo {author} {\bibfnamefont {S.~V.}\ \bibnamefont
			{Maleev}}, \bibinfo {author} {\bibfnamefont {V.~G.}\ \bibnamefont
			{Bar'yakhtar}},\ and\ \bibinfo {author} {\bibfnamefont {R.~A.}\ \bibnamefont
			{Suris}},\ }\bibfield  {title} {\bibinfo {title} {{The scattering of slow
				neutrons by complex magnetic structures}},\ }\href
	{https://www.osti.gov/biblio/4713671} {\bibfield  {journal} {\bibinfo
			{journal} {Soviet Phys.-Solid State (English Transl.)}\ }\textbf {\bibinfo
			{volume} {Vol: 4}} (\bibinfo {year} {1963})},\ \bibinfo {note} {institution:
		Ioffe Inst. of Physics, Leningrad}\BibitemShut {NoStop}%
	\bibitem [{\citenamefont {Duan}\ \emph {et~al.}(2015)\citenamefont {Duan},
		\citenamefont {Ren}, \citenamefont {Liu}, \citenamefont {Li}, \citenamefont
		{Liu},\ and\ \citenamefont {Zhang}}]{Duan2015}%
	\BibitemOpen
	\bibfield  {author} {\bibinfo {author} {\bibfnamefont {T.~F.}\ \bibnamefont
			{Duan}}, \bibinfo {author} {\bibfnamefont {W.~J.}\ \bibnamefont {Ren}},
		\bibinfo {author} {\bibfnamefont {W.~L.}\ \bibnamefont {Liu}}, \bibinfo
		{author} {\bibfnamefont {S.~J.}\ \bibnamefont {Li}}, \bibinfo {author}
		{\bibfnamefont {W.}~\bibnamefont {Liu}},\ and\ \bibinfo {author}
		{\bibfnamefont {Z.~D.}\ \bibnamefont {Zhang}},\ }\bibfield  {title} {\bibinfo
		{title} {{Magnetic anisotropy of single-crystalline Mn$_3$Sn in triangular
				and helix-phase states}},\ }\href {https://doi.org/10.1063/1.4929447}
	{\bibfield  {journal} {\bibinfo  {journal} {Appl. Phys. Lett.}\ }\textbf
		{\bibinfo {volume} {107}},\ \bibinfo {pages} {082403} (\bibinfo {year}
		{2015})}\BibitemShut {NoStop}%
	\bibitem [{\citenamefont {Tsai}\ \emph {et~al.}(2020)\citenamefont {Tsai},
		\citenamefont {Higo}, \citenamefont {Kondou}, \citenamefont {Nomoto},
		\citenamefont {Sakai}, \citenamefont {Kobayashi}, \citenamefont {Nakano},
		\citenamefont {Yakushiji}, \citenamefont {Arita}, \citenamefont {Miwa},
		\citenamefont {Otani},\ and\ \citenamefont {Nakatsuji}}]{Tsai2020}%
	\BibitemOpen
	\bibfield  {author} {\bibinfo {author} {\bibfnamefont {H.}~\bibnamefont
			{Tsai}}, \bibinfo {author} {\bibfnamefont {T.}~\bibnamefont {Higo}}, \bibinfo
		{author} {\bibfnamefont {K.}~\bibnamefont {Kondou}}, \bibinfo {author}
		{\bibfnamefont {T.}~\bibnamefont {Nomoto}}, \bibinfo {author} {\bibfnamefont
			{A.}~\bibnamefont {Sakai}}, \bibinfo {author} {\bibfnamefont
			{A.}~\bibnamefont {Kobayashi}}, \bibinfo {author} {\bibfnamefont
			{T.}~\bibnamefont {Nakano}}, \bibinfo {author} {\bibfnamefont
			{K.}~\bibnamefont {Yakushiji}}, \bibinfo {author} {\bibfnamefont
			{R.}~\bibnamefont {Arita}}, \bibinfo {author} {\bibfnamefont
			{S.}~\bibnamefont {Miwa}}, \bibinfo {author} {\bibfnamefont {Y.}~\bibnamefont
			{Otani}},\ and\ \bibinfo {author} {\bibfnamefont {S.}~\bibnamefont
			{Nakatsuji}},\ }\bibfield  {title} {\bibinfo {title} {Electrical manipulation
			of a topological antiferromagnetic state},\ }\href
	{https://doi.org/10.1038/s41586-020-2211-2} {\bibfield  {journal} {\bibinfo
			{journal} {Nature}\ }\textbf {\bibinfo {volume} {580}},\ \bibinfo {pages}
		{608} (\bibinfo {year} {2020})}\BibitemShut {NoStop}%
	\bibitem [{\citenamefont {Steinbrecher}\ \emph {et~al.}(2018)\citenamefont
		{Steinbrecher}, \citenamefont {Rausch}, \citenamefont {That}, \citenamefont
		{Hermenau}, \citenamefont {Khajetoorians}, \citenamefont {Potthoff},
		\citenamefont {Wiesendanger},\ and\ \citenamefont
		{Wiebe}}]{Steinbrecher2018}%
	\BibitemOpen
	\bibfield  {author} {\bibinfo {author} {\bibfnamefont {M.}~\bibnamefont
			{Steinbrecher}}, \bibinfo {author} {\bibfnamefont {R.}~\bibnamefont
			{Rausch}}, \bibinfo {author} {\bibfnamefont {K.~T.}\ \bibnamefont {That}},
		\bibinfo {author} {\bibfnamefont {J.}~\bibnamefont {Hermenau}}, \bibinfo
		{author} {\bibfnamefont {A.~A.}\ \bibnamefont {Khajetoorians}}, \bibinfo
		{author} {\bibfnamefont {M.}~\bibnamefont {Potthoff}}, \bibinfo {author}
		{\bibfnamefont {R.}~\bibnamefont {Wiesendanger}},\ and\ \bibinfo {author}
		{\bibfnamefont {J.}~\bibnamefont {Wiebe}},\ }\bibfield  {title} {\bibinfo
		{title} {Non-collinear spin states in bottom-up fabricated atomic chains},\
	}\bibfield  {journal} {\bibinfo  {journal} {Nat. Commun.}\ }\textbf {\bibinfo
		{volume} {9}},\ \href {https://doi.org/10.1038/s41467-018-05364-5}
	{10.1038/s41467-018-05364-5} (\bibinfo {year} {2018})\BibitemShut {NoStop}%
	\bibitem [{\citenamefont {Masuda}\ \emph {et~al.}(2024)\citenamefont {Masuda},
		\citenamefont {Seki}, \citenamefont {Ohe}, \citenamefont {Nii}, \citenamefont
		{Masuda}, \citenamefont {Takanashi},\ and\ \citenamefont
		{Onose}}]{Masuda2024}%
	\BibitemOpen
	\bibfield  {author} {\bibinfo {author} {\bibfnamefont {H.}~\bibnamefont
			{Masuda}}, \bibinfo {author} {\bibfnamefont {T.}~\bibnamefont {Seki}},
		\bibinfo {author} {\bibfnamefont {J.~I.}\ \bibnamefont {Ohe}}, \bibinfo
		{author} {\bibfnamefont {Y.}~\bibnamefont {Nii}}, \bibinfo {author}
		{\bibfnamefont {H.}~\bibnamefont {Masuda}}, \bibinfo {author} {\bibfnamefont
			{K.}~\bibnamefont {Takanashi}},\ and\ \bibinfo {author} {\bibfnamefont
			{Y.}~\bibnamefont {Onose}},\ }\bibfield  {title} {\bibinfo {title} {Room
			temperature chirality switching and detection in a helimagnetic mnau$_2$ thin
			film},\ }\bibfield  {journal} {\bibinfo  {journal} {Nat. Commun.}\ }\textbf
	{\bibinfo {volume} {15}},\ \href {https://doi.org/10.1038/s41467-024-46326-4}
	{10.1038/s41467-024-46326-4} (\bibinfo {year} {2024})\BibitemShut {NoStop}%
	\bibitem [{\citenamefont {Babkevich}\ \emph {et~al.}(2012)\citenamefont
		{Babkevich}, \citenamefont {Poole}, \citenamefont {Johnson}, \citenamefont
		{Roessli}, \citenamefont {Prabhakaran},\ and\ \citenamefont
		{Boothroyd}}]{Babkevich2012}%
	\BibitemOpen
	\bibfield  {author} {\bibinfo {author} {\bibfnamefont {P.}~\bibnamefont
			{Babkevich}}, \bibinfo {author} {\bibfnamefont {A.}~\bibnamefont {Poole}},
		\bibinfo {author} {\bibfnamefont {R.~D.}\ \bibnamefont {Johnson}}, \bibinfo
		{author} {\bibfnamefont {B.}~\bibnamefont {Roessli}}, \bibinfo {author}
		{\bibfnamefont {D.}~\bibnamefont {Prabhakaran}},\ and\ \bibinfo {author}
		{\bibfnamefont {A.~T.}\ \bibnamefont {Boothroyd}},\ }\bibfield  {title}
	{\bibinfo {title} {Electric field control of chiral magnetic domains in the
			high-temperature multiferroic cuo},\ }\href
	{https://doi.org/10.1103/PhysRevB.85.134428} {\bibfield  {journal} {\bibinfo
			{journal} {Phys. Rev. B}\ }\textbf {\bibinfo {volume} {85}},\ \bibinfo
		{pages} {134428} (\bibinfo {year} {2012})}\BibitemShut {NoStop}%
\end{thebibliography}
\end{document}